\documentclass[11pt]{article}
\usepackage{mathrsfs,amsmath,amsbsy,amssymb}

\def\be{\begin{equation}}
\def\ee{\end{equation}}
\def\bea{\begin{eqnarray}}
\def\eea{\end{eqnarray}}
\def\p{\partial}

\def\nn{\nonumber \\}

\title{Algebraically special axisymmetric solutions of
  the higher-dimensional vacuum Einstein equation} 

\author{Mahdi Godazgar and Harvey S. Reall\\Department of Applied Mathematics and Theoretical Physics \\ Centre for Mathematical Sciences \\ Wilberforce Road, Cambridge CB3 0WA, UK\\ mmg31@cam.ac.uk, hsr1000@cam.ac.uk}

\begin{document}

\maketitle

\begin{abstract}
A $d$-dimensional spacetime is ``axisymmetric'' if it possesses an $SO(d-2)$ isometry group whose orbits are $(d-3)$-spheres. In this paper, algebraically special, axisymmetric solutions of the higher dimensional vacuum Einstein equation (with cosmological constant) are investigated. Necessary and sufficient conditions for static axisymmetric solutions to belong to different algebraic classes are presented. Then general (possibly time-dependent) axisymmetric solutions are discussed. All axisymmetric solutions of algebraic types II, D, III and N are obtained.
\end{abstract}

\section{Introduction}

\subsection{Background}

A $d$-dimensional spacetime is ``axisymmetric'' if it possesses an $SO(d-2)$ isometry group whose orbits are $(d-3)$-spheres.  There are several motivations for studying axisymmetric solutions of the higher-dimensional vacuum Einstein equation (with cosmological constant)
\be
\label{eqn:einstein}
 R_{\mu\nu} = \Lambda g_{\mu\nu}.
\ee
These include the problem of finding an exact solution describing a black hole bound to a 3+1 dimensional brane in the (single brane) Randall-Sundrum model \cite{RS}, and determining the phase structure of General Relativity with a compactified dimension \cite{harmark}.

In $d=4$ dimensions, all static axisymmetric solutions of the
vacuum Einstein equation (with $\Lambda=0$) were obtained by Weyl, who showed that they
are characterized by a single axisymmetric harmonic function in $R^3$
(see e.g. \cite{wald}). Weyl's result has been generalized to higher
dimensions: the class of solutions of the $d$-dimensional vacuum
Einstein equation (with $\Lambda=0$) admitting $d-2$ commuting, orthogonal, non-null
Killing fields is specified by $d-3$ axisymmetric harmonic functions
in $R^3$ \cite{genweyl}. If one of the Killing fields generates time
translations and the others generate rotations then 
these solutions have isometry group $R \times
U(1)^{d-3}$, generalizing the $R \times U(1)$ symmetry of Weyl's
solutions. However, these solutions are not axisymmetric for $d>4$.

It is desirable to know the general static axisymmetric solution
in $d>4$ dimensions but, unfortunately, the Einstein equation
cannot be solved analytically for $d>4$ (or even for $d=4$ with $\Lambda \ne 0$). The impediment arises from the curvature of $S^{d-3}$ \cite{myers}. Note that $S^{d-3}$ is flat if
$d=4$, which is why the Einstein equation can be solved for $d=4$.

The goal of this paper is to determine whether the Einstein equation can be solved analytically if one makes the additional assumption that the spacetime is {\it algebraically special}. In $d=4$, an algebraically special static, axisymmetric spacetime must be type D (or O). For $\Lambda=0$, the only such solutions are, in Levi-Citiva's evocative terminology, the A-metrics, the
B-metrics, and the C-metric \cite{exactsolutionsbook}. The A-metrics are labelled by the
parameters $k\in\{-1,0,1\}$ and $M \ne 0$. The metric takes the generalized
Schwarzschild form
\be
 ds^2 = -U(r) dt^2 + \frac{dr^2}{U(r)} + r^2 d\Omega_k^2,
\ee
where $U(r) = k -2M/r$, and $d\Omega_k^2$ is the metric on
a space of  constant curvature with sign $k$. The B-metrics are analytic
continuations of the A-metrics in which the time coordinate $t$ is Wick
rotated to a spatial coordinate $\phi$ and $d\Omega_k^2$ to a Lorentzian metric of constant curvature $d\Sigma_k^2$:
\be
 ds^2 = U(r) d\phi^2 + \frac{dr^2}{U(r)} + r^2 d\Sigma_k^2.
 \ee
The C-metric describes a pair of black holes being accelerated apart by a conical singularity
\cite{kinnersleywalker}.

The first method for algebraic classification of the Weyl tensor in higher dimensions was proposed by de Smet \cite{desmetclass}. His spinorial method works only for the case $d=5$. We shall follow the alternative algebraic classification of Coley, Milson, Pravda and Pravdova (CMPP) \cite{algclass}, which applies for general $d$. This method uses ``Weyl aligned null directions'' (WANDs), which generalize the concept of principal null directions (PNDs) used for algebraic classification in $d=4$ dimensions. We shall explain the classification scheme in more detail below. For now, we just state that a $d>4$ spacetime need not admit a WAND, in which case it is algebraically general (type G), and that there is a notion of a ``multiple WAND'' that generalizes the 4d notion of a repeated PND. If a WAND exists, but not a multiple WAND, then the spacetime is type I. If a multiple WAND exists then the spacetime is type II, D, III, N or O. The conditions for these more special algebraic types will be described below. Finally, we note that WANDs need not be discrete for $d>4$, and the terminology PND is reserved for the case in which there are finitely many WANDs.

An important result in 4d is the Goldberg-Sachs theorem \cite{exactsolutionsbook}. This states that, for a solution of the Einstein equation (\ref{eqn:einstein}) that is not type O, a null vector field is a repeated PND if, and only if, it is tangent to a shear-free null geodesic congruence. This is not true in more than four dimensions \cite{frolov,bianchi}: a multiple WAND may not be geodesic, or it may be geodesic but shearing. An example of the former behaviour (with $\Lambda>0$) is the direct product $dS_3 \times S^{d-3}$ (for $d>4$), which is type D \cite{pravdatyped}. {\it Any} null vector field in $dS_3$ is a multiple WAND. Obviously not all such vector fields are geodesic. An example of the latter behaviour is the black string solution given by the product of the 4d Schwarzschild solution with a flat direction, which is also type D \cite{pravdatyped}. The multiple WANDs are the repeated PNDs of the Schwarzschild solution. These are geodesic (by the Goldberg-Sachs theorem). The associated congruences of geodesics expand in the
Schwarzschild directions but not in the flat direction, hence they are shearing.\footnote{
See section 5.4 of Ref. \cite{kerrschild} for discussion of constraints on the shear of a geodesic multiple WAND.}

A partial generalization of the Goldberg-Sachs theorem does hold for vacuum spacetimes of type III or type N, for which it has been shown that the multiple WAND must be geodesic \cite{bianchi}.\footnote{\label{note:gs} The argument of Ref. \cite{bianchi} assumes $\Lambda=0$ but it is straightforward to generalize to $\Lambda \ne 0$.}

\subsection{Summary of results}

In this paper, we start (in section \ref{sec:static}) by considering static, axisymmetric solutions and determine the condition for them to be algebraically special. It was shown in Ref. \cite{pravdatyped} that a static solution must be of algebraic type G, I, D or O. We derive simple necessary and sufficient conditions for a solution to belong to the various algebraic types. We also show that many analytic solutions are type G in one open subset of the spacetime and type I in another. This suggests that distinguishing between type G and type I solutions is not very useful in practice, and that the type I condition alone will not be much help in finding new solutions (just as in 4d, where type I is algebraically general).\footnote{This also suggests that it might be convenient to change nomenclature and reserve the term "algebraically special" for spacetimes admitting a multiple WAND, just as in 4d.}

In the rest of the paper, we relax the condition of staticity and consider general (possibly time-dependent) algebraically special axisymmetric solutions. The starting point of our analysis is the observation that,  for $d>4$, the action of $SO(d-2)$ on $S^{d-3}$ must be {\it orthogonally transitive} \cite{exactsolutionsbook}, i.e., spacetime is locally a product $M_3 \times S^{d-3}$ with warped product metric
\be
 \label{eqn:orthogonal}
 ds^2 = g_{ab}(x) dx^a dx^b + E(x)^2 d\Omega^2,
\ee
for some 3-metric $g_{ab}$ and function $E(x)$ on $M_3$, where $d\Omega^2$ is the metric on $S^{d-3}$ normalized to unit radius. The analysis naturally divides into two cases depending on whether or not the WAND is axisymmetric, i.e., invariant under $SO(d-2)$. 

In section \ref{sec:axiwand}, we consider the case in which the WAND is axisymmetric. In this case, there is a partial generalization of the Goldberg-Sachs theorem. We show that the only axisymmetric solution of type II or D with a {\it non-geodesic} axisymmetric multiple WAND is $dS_3 \times S^{d-3}$ with $\Lambda>0$. For any other axisymmetric type II or D vacuum solution, an axisymmetric multiple WAND must be geodesic. As already noted, this is also true for type III and type N \cite{bianchi}. (Hence any axisymmetric vacuum solution with an axisymmetric but non-geodesic multiple WAND must be type O, or $dS_3 \times S^{d-3}$.) The axisymmetry implies that the null geodesic congruence tangent to the WAND has vanishing rotation, hence it is hypersurface-orthogonal. We determine all solutions with an axisymmetric geodesic WAND without assuming a particular algebraic type. The solutions are all type II or more special.\footnote{
For $\Lambda=0$, this implies that these solutions belong to the class of spacetimes discussed in Ref. \cite{nontwisting}, i.e., those admitting a hypersurface orthogonal geodesic multiple WAND. The dependence on the affine parameter along the geodesics was determined in that paper.} There are several classes.

\begin{itemize}
\item
 Type O (conformally flat) solutions. Irrespective of axisymmetry, the only such solutions are Minkowski, de Sitter, and anti-de Sitter spacetimes.

\item
The Schwarzschild solution, generalized to allow for flat or hyperbolic slices (i.e.,  higher-dimensional analogues of the A-metrics) and a cosmological constant. The metric is given by equation \eqref{eqn:schw}. The solution is type D. The null congruence tangent to the WAND has vanishing shear and non-vanishing expansion, so these solutions are a subset of the higher-dimensional Robinson-Trautman family of solutions (defined to be solutions admitting a null geodesic congruence with vanishing shear and rotation and non-vanishing expansion) obtained in Ref. \cite{ortaggio}.

\item
``Black string'' solutions obtained, for $\Lambda=0$, by foliating Minkowski spacetime with $(d-1)$-dimensional Minkowski, or de Sitter, slices and replacing the slices with a Schwarzschild, or Schwarzschild-de Sitter, metric respectively. In the former case, this gives the familiar Schwarzschild black string solution. There is an analagous construction for $\Lambda>0$ based on a de Sitter foliation of de Sitter spacetime, and for $\Lambda<0$ based on Minkowski, de Sitter, or anti-de Sitter foliations of anti-de Sitter spacetime. The latter includes the anti-de Sitter black string of Ref. \cite{chamblin}. The metric of these solutions is given by equation \eqref{eqn:string}. They are all type D. The null congruence associated with the WAND has non-vanishing expansion and shear.

\item
For $\Lambda>0$, $dS_3 \times S^{d-3}$ is type D. As discussed above, a {\it general} multiple WAND of this spacetime is non-geodesic. However, any null geodesic congruence in $dS_3$ defines a geodesic multiple WAND, which is why this solution is mentioned here. Such a congruence may be expanding and shearing or non-expanding and non-shearing (in the latter case the solution is a special case of the Kundt solutions discussed next).

\item
Axisymmetric Kundt solutions (section \ref{sec:kundt}). A Kundt spacetime is a spacetime admitting a null geodesic congruence with vanishing expansion, rotation and shear \cite{exactsolutionsbook}. Such solutions are type II, or more special, for any $d \ge 4$ \cite{ricci}. In general they involve arbitrary functions of time.  In our axisymmetric case, these solutions are expressed in terms of solutions of certain ODEs that cannot be solved analytically in general. We show that some of these solutions are type D or N (but not III). The type N solutions can be obtained analytically. They describe gravitational waves in Minkowski (eq. (\ref{eqn:ppwave})), de Sitter (eq. (\ref{eqn:dswave})) or anti-de Sitter (eq. (\ref{eqn:adswave})) spacetime. The general type D solution is cohomogeneity-1 with 
surfaces of homogeneity $M_2 \times S^{d-3}$ where $M_2$ is 2d Minkowski or (anti-) de Sitter spacetime:
\be
\label{eqn:typeDkundt}
 ds^2 = dz^2 + A(z)^2 d\Sigma^2 + R(z)^2 d\Omega^2,
\ee
where $d\Sigma^2$ is the metric on $M_2$. The functions $A(z)$ and $R(z)$ can be determined analytically only in special cases e.g. the product spaces $dS_3 \times S^{d-3}$, $dS_2 \times S^{d-2}$ and $AdS_2 \times H^{d-2}$, or for flat $M_2$ with $\Lambda=0$. Various solutions of this form have been discussed previously in the literature. For positively curved $M_2$, if one analytically continues to Riemannian signature (so that $M_2$ becomes $S^2$) then these are metrics of the form discussed by B\"ohm. He proved that, for low enough $d>4$, and $\Lambda>0$, there exist infinitely many Einstein metrics on spheres and products of spheres of the form \eqref{eqn:typeDkundt} \cite{bohm1}. for $\Lambda \le 0$ he constructed complete, non-compact metrics of the form \eqref{eqn:typeDkundt} \cite{bohm2}. The Lorentzian interpretation of some of the B\"ohm solutions has been discussed in Ref. \cite{gibbons}. Some singular solutions with flat $M_2$ and $\Lambda=0$ were discussed in Ref. \cite{gregory}, analagous solutions with non-flat $M_2$ were discussed in Ref. \cite{gregory2}. A regular solution with $d=5$, flat $M_2$ and $\Lambda<0$ describes the metric dual to the ground state of ${\cal N}=4$ super Yang-Mills theory on $R \times S^1 \times S^2$ (with fermions periodic on $S^1$) \cite{horowitz}. A solution with $d=5$, negatively curved $M_2$ and $\Lambda<0$ describes the bulk near-horizon geometry of an extremal charged Randall-Sundrum black hole, or the metric dual to the ground state of ${\cal N}=4$ SYM in $AdS_2 \times S^2$ \cite{kaus}. More generally, solutions of the form \eqref{eqn:typeDkundt} with $\Lambda<0$ that are asymptotically locally anti-de Sitter presumably describe the metrics dual to the ground state of a CFT in $M_2 \times S^{d-3}$. 

For $d=4$, it was proposed in Ref. \cite{ortaggio2} that some type II Kundt solutions describe gravitational waves propagating in a ``background'' spacetime described by a type D Kundt solution. The same is true for $d>4$: given a type D background of the form \eqref{eqn:typeDkundt}, one can construct explicitly axisymmetric type II Kundt solutions which describe gravitational waves propagating along the space $M_2$ of this background. (Some particular examples of such solutions were obtained in Refs. \cite{jakimowicz1,jakimowicz2}.) For $\Lambda<0$, some of these solutions will be asymptotically locally AdS, and will describe metrics dual to certain CFT states in $M_2 \times S^{d-3}$ for which there is a null energy-momentum flux along $M_2$. 

\end{itemize}

The second case to consider is when the WAND is not axisymmetric. Acting with $SO(d-2)$ generates a continuously infinite family of WANDs, which suggests that the solution should have an enhanced symmetry. This is indeed the case: assuming that the solution admits a multiple WAND, we are able to show that $SO(d-2)$ is enhanced to the de Sitter symmetry $SO(1,d-2)$, with $dS_{d-2}$ orbits, and the 
only non-trivial (i.e. not type O) solutions are:
\begin{itemize}
\item
For any $\Lambda$, a Kaluza-Klein bubble solution \cite{witten} (i.e. a higher-dimensional analogue of the B-metrics) obtained by analytic continuation of the Schwarzschild solution (so that $S^{d-2} \rightarrow dS_{d-2}$). The metric is given by equation \eqref{eqn:bubble}. This is type D.

\item
For positive $\Lambda$ there is a $dS_{d-2} \times S^2$ solution. This is also type D.
\end{itemize}
For both solutions, any null vector field tangential to the $dS_{d-2}$ orbits of $SO(1,d-2)$ is a multiple WAND. Hence, these are further examples of type D vacuum solutions for which multiple WANDs need not be geodesic.\footnote{
This answers a question posed in \cite{pravdatyped}: do there exist $d=5$, $\Lambda=0$, type D solutions with non-geodesic multiple WAND? For the type D examples with non-geodesic multiple WANDs encountered in our analysis, the WANDs are not discrete so they are not PNDs. So one could restate the question as: does there exist a $d=5$ vacuum type D spacetime with a non-geodesic repeated PND?}

Combining these results, we learn that any algebraically special axisymmetric solution that is {\it not} encompassed by our analysis must be type I and such that every WAND is either not invariant under $SO(d-2)$ or is invariant under $SO(d-2)$ but is not geodesic. In particular, any axisymmetric solution admitting a multiple WAND is one of the solutions listed above. It is convenient to summarize our results according to algebraic type:
\begin{itemize}
\item
Type O: the only type O Einstein solutions are Minkowski or (anti)-de Sitter spacetime.
\item
Type N: the only axisymmetric solutions are the type N axisymmetric Kundt solutions.
\item
Type III: there are no axisymmetric type III solutions.
\item
Type D: all axisymmetric solutions are contained in the following list: Kaluza-Klein bubble; $dS_{d-2} \times S^2$; generalized Schwarzschild; generalized black string; solutions of the form (\ref{eqn:typeDkundt}).
\item
Type II: the only axisymmetric solutions are the type II axisymmetric Kundt solutions.
\item
Type I: if a WAND of an axisymmetric type I solution is axisymmetric then it is non-geodesic.
\end{itemize}
Note that the type D solutions all have isometry groups larger than the $SO(d-2)$ that was assumed initially.

We can compare these results to those of de Smet, who classified {\it static}, axisymmetric $d=5$ spacetimes belonging to classes 22 and \underline{22} in his classification scheme for $\Lambda=0$ \cite{desmetclass} and $\Lambda \ne 0$ \cite{desmet22}. For $d=5$, our list of type D solutions is very similar to the set of the solutions that he found.\footnote{We are taking results from the arXiv version of Ref. \cite{desmetclass}, which differs significantly from the published version. De Smet worked in Euclidean signature and hence could not distinguish between the Schwarzschild solution and a KK bubble.} One significant difference is for $\Lambda=0$, where he found a solution that is not on our list (eq. 4.6 of Ref. \cite{desmetclass}, a ``homogeneous wrapped object''). The results of section \ref{sec:static} below show that this solution is type G in the CMPP classification. Curiously, no analagous solution with $\Lambda \ne 0$ was obtained in Ref. \cite{desmet22}. Some of the generalized ``black string'' solutions that we found (eq. \eqref{eqn:string}) do not appear in de Smet's results. It would be interesting to understand the connections between the de Smet scheme and the CMPP scheme.

It is also interesting to compare our results with results for $d=4$. For $d=4$, axisymmetry is much less restrictive than for $d>4$. This is because, the action of a 1-dimensional group such as $SO(2)$ need not be orthogonally transitive, so the orthogonal decomposition (\ref{eqn:orthogonal}) is not always possible, e.g. it does not apply for the Kerr solution. The natural $d=4$ analogues of $d>4$ axisymmetric spacetimes are spacetimes with a spacelike, hypersurface-orthogonal Killing vector field. Coordinates can then be chosen so that the Killing field is $\partial/\partial \phi$ and there is a discrete isometry $\phi \rightarrow -\phi$. The metric then can be written in the form (\ref{eqn:orthogonal}) with $d\Omega^2 = d\phi^2$. So this is the class of $d=4$ spacetimes analagous to our spacetimes. Algebraically special $d=4$ vacuum solutions (with $\Lambda=0$) with these symmetries were first classified by Kramer and Neugebauer \cite{KN}, so we shall refer to them as KN solutions.

By the Goldberg-Sachs theorem, the null congruence tangent to the repeated PND is geodesic and shear-free. For KN solutions it can be shown that it is also rotation-free, i.e., hypersurface-orthogonal \cite{KN}. Hence a KN solution belongs to the Robinson-Trautman (RT) or Kundt family of solutions depending on whether the congruence associated to the repeated PND is expanding or not. In 4d, the general vacuum solution belonging to either of these classes involves arbitrary functions of time, and cannot be written down in closed form \cite{exactsolutionsbook}. However, with the KN symmetries, the general Kundt (but not RT) solution can be obtained in closed form \cite{KN}. Special cases of RT solutions with the KN symmetries are the A-metrics and the C-metric. The Kundt family of solutions contains the B-metrics.\footnote{We have ignored a special case that arises in the KN analysis, which occurs when the repeated PND is not invariant under the discrete symmetry (and must therefore be mapped to another repeated PND so the spacetime is type D). KN showed that the only such solution is the $k=0$ B-metric. However, this admits a second hypersurface-orthogonal spacelike Killing field, and the associated discrete symmetry {\it does} preserve the repeated PNDs. This implies that the solution also belongs to the Kundt class.}

The main differences betwen these $d=4$ results and our results are (i) the absence of time-dependent axisymmetric $d>4$ RT solutions; (ii) the absence of a $d>4$ analogue of the C-metric. The first difference extends beyond axisymmetry: the general class of $d>4$ RT solutions was investigated in Ref. \cite{ortaggio} and found to be considerably simpler than the $d=4$ class. In particular, the only $d>4$ RT solutions with non-vanishing ``mass function'' are simple static generalizations of the Schwarzschild solution, in contrast with the $d=4$ case where such solutions are generically time-dependent. Concerning point (ii), to explain what we mean by ``analogue'', we note that the main interest in constructing such a solution is to obtain an exact solution describing a black hole on a Randall-Sundrum brane, as explained in Ref. \cite{EHM}. Such a solution would have $d=5$, $\Lambda<0$ and would be axisymmetric (if the black hole were spherically symmetric on the brane) and describe an object with an event horizon accelerating along the axis of symmetry. The $d=4$ C-metric belongs to the Weyl class, the RT class and the class of type D metrics. However, no $d>4$ analogue was found in the generalized Weyl class (for $\Lambda=0$) \cite{genweyl} or, as we have just discussed, the $d>4$ RT class \cite{ortaggio}. Our results demonstrate that this negative conclusion extends to the $d>4$ type D class too. However, we note that a type D metric of the form \eqref{eqn:typeDkundt} does describe the {\it near-horizon geometry} of an extremal charged Randall-Sundrum black hole \cite{kaus}.

\subsection{Algebraic classification in higher dimensions}

\begin{center}
\begin{table}
\begin{tabular}{ll}
type I & $C_{0i0j}=0$ \\
type II & $C_{0i0j}=C_{0ijk}=0$ \\
type D & $C_{0i0j}=C_{0ijk} = C_{1i1j} = C_{1ijk} = 0$ \\
type III & $C_{0i0j} = C_{0ijk} = C_{ijkl}=C_{01ij}=0$ \\
type N & $C_{0i0j} = C_{0ijk} = C_{ijkl}=C_{01ij}=C_{1ijk}=0 $ \\
type O & $C_{\alpha\beta\gamma\delta}=0$
\end{tabular}
\caption{Conditions for the various algebraic types. Note that the definition of type D involves secondary classification, i.e., consideration of $n$ as well as $\ell$.}
\label{table:classification}
\end{table}
\end{center}

For convenience, we shall review briefly the CMPP classification scheme \cite{algclass}. This involves a null basis $e_0 \equiv \ell$, $e_1 \equiv n$, $e_i=m_i$, $i=2 \ldots d-1$ where $\ell$ and $n$ are null vectors obeying $\ell \cdot n = 1$, and $m_i$ are an orthonormal set of spacelike vectors orthogonal to $\ell$ and $n$. Consider a new frame related by a boost
\be
 \hat{\ell} = \lambda \ell, \qquad \hat{n} = \lambda^{-1} n. \qquad \hat{m}_i = m_i.
\ee
A covariant tensor component $T_{a_1 \ldots a_p}$ is said to be of {\it boost weight} $s$ if its value in the new basis is related to its value in the old basis by
\be
 T_{\hat{a}_1 \ldots \hat{a}_p} = \lambda^s T_{a_1 \ldots a_p}.
\ee
We are primarily interested in the Weyl tensor. The components of highest boost weight are the components $C_{0i0j}$, which have $s=2$. The null direction $\ell$ is called a {\it Weyl aligned null direction} (WAND) if, and only if, these components vanish. This is independent of how the other vectors of the basis are chosen. For $d=4$, WANDs are the same as PNDs. However, for $d>4$, in general, no WAND exists and the spacetime is called type G. If a WAND does exist then the solution is called algebraically special. Another important difference in higher dimensions is that WANDs need not be discrete. The terminology PND is reserved for the case in which there are finitely many WANDs.

If $\ell$ satisfies the condition that all Weyl tensor components of boost weight $2$ and $1$ vanish then $\ell$ is a {\it multiple WAND}. Using the tracefree property of the Weyl tensor, the multiple WAND condition is $C_{0i0j}=C_{0ijk}=0$. If an algebraically special spacetime does {\it not} admit a multiple WAND then it is called type I. If it does admit a multiple WAND then it is type II, or perhaps more special. The more special types are: type III if all Weyl components of boost weight $2,1,0$ vanish, type N if all components of boost weight $2,1,0,-1$ vanish and type O if the Weyl tensor vanishes. The explicit conditions for these more special algebraic types are given in table \ref{table:classification}. The algebraic types are mutually exclusive, i.e., a spacetime of type III is not also type II.

So far, we have discussed only the WAND $\ell$, which gives the so-called {\it primary} classification. One can then perform a more refined {\it secondary} classification by examining whether, or not, for $\ell$ given by the primary classification, it is possible to choose $n$ to make further Weyl tensor components vanish. For example, if spacetime is type I with WAND $\ell$ and it is possible to choose $n$ so that $C_{1i1j}=0$ then the spacetime is said to be type I$_i$. We shall not make use of secondary classification {\it except} in defining type D spacetimes. A spacetime of primary type II, with multiple WAND $\ell$, is said to be type D if $n$ can be chosen so that $C_{1i1j}=C_{1ijk}=0$. Note that this implies that $n$ also is a multiple WAND.

We are interested in solutions of the vacuum Einstein equation (\ref{eqn:einstein}). For such spacetimes, the WAND conditions can be reformulated in terms of the Riemann tensor:
\be
\label{eqn:WANDcondition} 
R_{0i0j}=0 \Leftrightarrow {\rm WAND}, \qquad R_{0i0j}=R_{0ijk}=0 \Leftrightarrow {\rm multiple \, WAND}.
\ee
We shall make use of several general results for warped product spacetimes. A warped product is a spacetime of the form
\be
 ds^2 = A(y)^2 g_{ab}(x) dx^a dx^b + B(x)^2 g_{ij}(y) dy^i dy^j,
\ee
where $g_{ab}$ is Lorentzian and $g_{ij}$ is Riemannian. Such a spacetime is type D or O if the Lorentzian factor is (i) two-dimensional; (ii) a three-dimensional Einstein space; (iii) a type D Einstein space \cite{pravdatyped}.

\section{Static, axisymmetric, solutions}

\label{sec:static}

In this section we consider higher-dimensional solutions that are static and axisymmetric, i.e., they admit a hypersurface orthogonal timelike Killing vector field that commutes with the generators of $SO(d-2)$. Introduce coordinates adapted to the isometries:
\be
 ds^2 = -A(r,z)^2 dt^2 + B(r,z)^2 (dr^2 + dz^2) + C(r,z)^2 d\Omega^2.
\ee
The components of the Einstein equation (\ref{eqn:einstein}) for this metric are given in Ref.
\cite{charmousis}. Define a complex coordinate $ w \equiv (r+iz) /\sqrt{2}$.
Consider $R^r_r - R^z_z + 2iR^r_z=0$. This gives
\be
 \frac{\partial^2 A}{A} -2\frac{\partial A \partial B}{A B} +(d-3)\left( \frac{\partial^2 C}{C} -2\frac{\partial B \partial C}{B C}\right) =0,
\ee
where $\p \equiv \p/\p w$. This implies
\begin{equation}
\label{eqn:Beq1}
 \frac{\partial B}{B}=\frac{C^{d-3} \partial^2 A+(d-3) A C^{d-4} \partial^2 C}{2\partial (A C^{d-3})}.
\end{equation}
We must consider the denominator since it could vanish identically, i.e., $AC^{d-3}$ might be constant. The equation $R^t_t+(d-3)R^{\theta_1}_{\theta_1}=(d-2) \Lambda$ implies\footnote{
We have defined $\nabla = (\p_r,\p_z)$, $\Delta = \nabla^2$, and indices are raised with the flat metric $dr^2+dz^2$.}
\begin{equation}
 \frac{\Delta (A C^{d-3})}{A C^{d-3}} = \dfrac{B^2}{C^2} \left[(d-3)(d-4)-(d-2) \Lambda C^2 \right]. \label{eqndel}
\end{equation}
Constancy of $\partial (A C^{d-3})$ implies the RHS must vanish hence $C$ is a constant and $\Lambda$ is positive. But then $A$ must also be constant so the spacetime has a flat time direction. This is incompatible with positive $\Lambda$. Hence $AC^{d-3}$ cannot be constant.

If $A$ and $C$ are known then equation (\ref{eqn:Beq1}) determines $B$. 
Furthermore, $R^r_r + R^z_z - 2R^t_t=0$ implies
\begin{equation}
 \frac{(\nabla B)^2}{B^2}-\frac{\Delta B}{B} = \frac{(d-3) \Delta C}{2 C}-\frac{\Delta A}{2 A}- (d-3) \frac{ \nabla A \cdot \nabla C}{A C}. \label{eqn:Beq2}
\end{equation}
It can be checked that this is compatible with equation (\ref{eqn:Beq1}).

We assume that the spacetime is algebraically special, so it admits a WAND $\ell$. We shall assume for now that the WAND is axisymmetric. Assuming $d>4$, this implies that it is orthogonal to $S^{d-3}$. By rescaling $\ell$ we can arrange that 
\be
 \ell = \frac{\partial}{\partial t} + \frac{A}{B} \left( \cos \alpha(r,z) \frac{\partial}{\partial r} + \sin \alpha(r,z) \frac{\partial}{\partial z} \right),
\ee
for some function $\alpha(r,z)$. Staticity implies that
\be
 n = \frac{\partial}{\partial t} - \frac{A}{B} \left( \cos \alpha(r,z) \frac{\partial}{\partial r} + \sin \alpha(r,z) \frac{\partial}{\partial z} \right),
\ee
is also a WAND\footnote{
These null vectors don't obey $\ell \cdot n=1$ but this can arranged by rescaling them, which doesn't affect anything below.}, i.e., WANDs come in pairs, which implies that the algebraic type must be I, D or O \cite{pravdatyped}. Choose
\be
 m^2 = \frac{1}{B} \left( -\sin \alpha(r,z) \frac{\partial}{\partial r} + \cos \alpha (r,z) \frac{\partial}{\partial z} \right),
\ee
and 
\be
 m^\alpha = \frac{1}{C} \hat{e}^\alpha, \qquad \alpha = 3, \ldots, d-1
\ee
where $\hat{e}^\alpha$ is a vielbein for $S^{d-3}$. We find that the WAND condition \eqref{eqn:WANDcondition} reduces to
\be
 {\rm Re} \left( e^{2i\alpha} W \right) = X,
\ee
where
\begin{equation}
 W=\frac{\partial^2 A}{A} -\frac{\partial^2 C}{C} -2\frac{\partial A \partial B}{A B}+2\frac{\partial B \partial C}{B C}=(d-2)  \frac{ C^{d-4}\left( \partial^2 A \partial C-\partial A \partial^2 C \right)}{\partial(A C^{d-3})},
\end{equation}
\begin{equation}
 X=\frac{\Delta A}{2 A}-\frac{\Delta B}{B}+\frac{\Delta C}{2 C}-\frac{\nabla A \cdot \nabla C}{A C}+\frac{(\nabla B)^2}{B^2}=\frac{(d-2) A^2}{2 C} \nabla \cdot (\frac{\nabla C}{A^2}),
\end{equation}
where the second equality in each case follows from equations (\ref{eqn:Beq1}) and (\ref{eqn:Beq2}).
The spacetime is algebraically special if, and only if, there
exists a real solution $\alpha$ of the WAND condition. Hence
\be
 |W| \ge |X| \qquad \Leftrightarrow \qquad \rm{algebraically \; special}
\ee
Now consider the additional condition required for a multiple WAND (equation \eqref{eqn:WANDcondition}).  This gives the single equation ${\rm Im} \left( e^{2i\alpha} W \right)=0$. Combining with the type I condition gives
\be
 e^{2 i \alpha} W = X.
\ee
We conclude that, assuming an axisymmetric WAND
\bea
\label{eqn:staticclass}
 |W| &<& |X| \qquad \Leftrightarrow \qquad {\rm Type \; G} \nn
 |W| &>& |X| \qquad \Leftrightarrow \qquad {\rm Type \; I} \\ 
|W| &=& |X|  \qquad \Leftrightarrow \qquad {\rm Type \; D \; or \; O} \nonumber
\eea We shall now comment on our assumption that the WAND is axisymmetric. In general, this need not be true. However. for odd $d$, $S^{d-3}$ is even dimensional so the projection of $\ell$ onto the sphere must vanish somewhere. Working at such a point, we can argue as above to arrive at equations (\ref{eqn:staticclass}) that depend only on $r$ and $z$ and must therefore 
hold {\it everywhere} on the sphere, which implies the existence of an axisymmetric WAND. Hence, for odd $d$, there is no loss of generality in restricting 
to an axisymmetric WAND. For even $d$, this argument does not work. 
However, in section \ref{sec:WANDnotaxi}, we shall consider axisymmetric spacetimes with a 
non-axisymmetric {\it multiple} WAND, and show that no such spacetime is 
static and axisymmetric.\footnote{Actually, the spacetimes we 
find there are ``static'' and ``axisymmetric'' but they are not ``static 
and axisymmetric'' because the generator of time translations does not 
commute with the generators of axisymmetry.} Therefore the multiple WAND of a static axisymmetric type D spacetime must be axisymmetric. However, it is possible that some spacetimes with even $d>4$ and $|W|<|X|$ may be type I with a non-axisymmetric WAND.

To illustrate these conditions, consider the $z$-independent $d=5$, $\Lambda=0$ solution of Ref. \cite{chodosetal}, written in the form given in Ref. \cite{gross}
\bea
 ds^2 &=& - \left( \frac{1-m/R}{1+m/R} \right)^{2/\alpha} dt^2 + \left( \frac{1-m/R}{1+m/R} \right)^{2\beta /\alpha} dz^2 \nn &+& \left( 1+ \frac{m}{R} \right)^4  \left( \frac{1-m/R}{1+m/R} \right)^{2(\alpha-\beta-1)/\alpha} \left( dR^2 + R^2 d\Omega^2 \right),
\eea
where $\alpha = \sqrt{\beta^2 + \beta +1}$. Assume $m \ne 0$ (so the spacetime is not flat). Then a calculation reveals that $|W|=|X|$ if, and only if, $\beta=0$ or $\beta=1$. The first possibility gives the Schwarzschild black string. The second possibility gives a boost invariant singular spacetime discussed in Ref. \cite{gregory}. This spacetime is of the form \eqref{eqn:typeDkundt} discussed in the introduction.

Another interesting example is the static Kaluza-Klein bubble (the product of a flat time direction with the Euclidean Schwarzschild solution). This can be obtained by taking the limit $\beta \rightarrow \infty$ of the above metric. This spacetime has $W=0$, $X\ne 0$ hence it is type G.

Since type G is distinguished from type I only by an inequality, it is possible that there exist (connected) analytic spacetimes that are type G in some open subset of spacetime and type I in some other open subset. Indeed, if we choose $m>0$ and $\beta = 1/2$ in the above metric then it is type G for $R \sim m$ but type I for $R \gg m$. \footnote{The general behaviour appears to be that, for $0< \beta <1$ and $m>0$, the solution is type G near $R=m$ and type I for $R \gg m$. For (finite) $\beta>1$ and $m>0$, it is type I near $R=m$ and type G for $R \gg m$.}
As discussed in the introduction, this kind of behaviour suggests that the type I condition alone will not be much help in solving the Einstein equation.

It would be nice to use the type D condition obtained above to solve the Einstein equation. However, we have not made progress using the coordinates employed here. (Even in $d=4$, this approach would not work for $\Lambda \ne 0$.) However, in subsequent sections we shall see that all static axisymmetric type D solutions {\it can} be found, indeed we shall relax the condition of stationarity and determine all axisymmetric type D solutions.

\section{Axisymmetric solutions with an axisymmetric WAND}

\label{sec:axiwand}

\subsection{Introducing coordinates}

In this section we shall consider general (possibly time-dependent) axisymmetric spacetimes with an axisymmetric WAND. First we shall introduce coordinates adapted to the WAND. Consider the metric in the form \eqref{eqn:orthogonal}, where spacetime is locally a warped product $M_3 \times S^{d-3}$. Axisymmetry implies that the WAND is tangential to $M_3$.

We shall choose the local coordinates $x^a$ on $M_3$ as follows. Pick a 2-surface in $M_3$ transverse to the WAND $\ell$ and let $x^{\hat{a}}$ be coordinates on this surface, where $\hat{a}=1,2$. Now carry these coordinates to the rest of spacetime along the integral curves of $\ell$, and let $\hat{r}$ be the parameter distance along these curves. Now use $(x^{\hat{a}},\hat{r})$ as coordinates on $M_3$, so $\ell=\partial/\partial \hat{r}$. The metric takes the form
\be
 ds^2 = 2g_{\hat{r} \hat{a}} d\hat{r} dx^{\hat{a}} + g_{\hat{a} \hat{b}} dx^{\hat{a}} dx^{\hat{b}} + E^2 d\Omega^2,
\ee
where $g_{r\hat{a}} \ne 0$ for some $\hat{a}$. Without loss of generality we may assume $g_{\hat{r} 1} \ne 0$. Now let $r=\int g_{\hat{r} 1} (\hat{r}, x^{\hat{a}}) d\hat{r}$, $v=x^1$, $z=x^2$ and use coordinates $(v,r,z)$. In this chart, the metric takes the form
\bea
ds^2 &=& 2 \left[dv+B(v,r,z) dz\right]\left[dr - \frac{1}{2} U(v,r,z) \left(dv
  +B(v,r,z)dz\right) +C(v,r,z) dz \right] \nn &+& D(v,r,z)^2 dz^2 + E(v,r,z)^2
  d\Omega^2,
 \eea
for some functions $U,B,C,D,E$. The WAND is proportional to $\partial/\partial r$ so we can rescale it so that $\ell=\partial/\partial r$. It is convenient to complete this to a null basis as follows:
\bea
\label{eqn:basis}
 \ell_a dx^a &=& dv + B dz, \qquad n_a dx^a = dr - \frac{1}{2} U
 \ell_a dx^a +C dz, \nn  m^2 &=& D dz, \qquad m^\alpha =
 E \hat{e}^\alpha,
\eea
where $\hat{e}^\alpha$ ($\alpha = 3 \ldots d-1$) is an orthonormal basis of 1-forms on $S^{d-3}$ with no dependence on $v,r,z$. We shall denote the spacelike basis 1-forms collectively as $m^i$,
$i=2\ldots d-1$.

Now consider the null congruence associated with the WAND. This is geodesic if, and only if, $\partial_r B \equiv 0$. The ``expansion matrix'' of the congruence is 
\be
\label{eqn:expmatrix}
 S_{ij} \equiv m_i^\mu m_j^\nu \nabla_{(\mu}\ell_{\nu)}=\textup{diag}\left(\frac{\partial_r D}{D},\frac{\partial_r E}{E},\ldots, \frac{ \partial_r E}{E}\right).
\ee
The expansion of the congruence is the trace of this matrix and the shear tensor is the traceless part. The rotation matrix of the congruence vanishes: this is a consequence of axisymmetry.

\subsection{Type II or D implies geodesic or $dS_3 \times S^{d-3}$} 

\label{Dimpliesgeo}

Consider the case that $\ell$ is a {\it multiple} WAND. It has been shown that for a type III or N Einstein spacetime, the multiple WAND is geodesic\footnote{See footnote \ref{note:gs}.} \cite{bianchi}. In this section we shall prove a generalization of this result for axisymmetric vacuum spacetimes with an axisymmetric multiple WAND, of principal type II, i.e., the spacetime is type II or D.

Assume that spacetime is type II or D and that the multiple WAND is axisymmetric and {\it not} geodesic. Define the matrix
\be
 \Phi_{ij} = C_{0i1j}.
\ee
Let $\Phi = \Phi_{ii}$. Define also
\be
 L_i = m_i^\mu \ell^\nu \nabla_\nu \ell_\mu,
\ee
so the WAND is geodesic if, and only if, $L_i = 0$.  It has been shown \cite{pravdatyped} that if the multiple WAND is {\it not} geodesic then $\Phi_{ij}$ must be symmetric with an eigenvalue equal to $-\Phi$, with associated eigenvector $L_i$. In our case, axisymmetry implies that $L_\alpha =0$. Hence we must have $\Phi_{22} = -\Phi$. Now axisymmetry implies that $\Phi_{2\alpha}=0$ and $\Phi_{\alpha \beta} \propto \delta_{\alpha\beta}$, so we must have
\be
 \Phi_{ij} = \Phi \, {\rm diag} \left( -1, 2/(d-3), \ldots, 2/(d-3) \right).
\ee 
For any type D spacetime satisfying \eqref{eqn:einstein}, equation (27) of Ref. \cite{pravdatyped} relates $\Phi_{ij}$ to $S_{ij}$:
\be
0=-S_{kk} \Phi_{(ij)} + \Phi S_{ij} + \Phi_{(ik)}S_{kj} + \Phi_{(jk)} S_{ki} + 3 \left( \Phi_{[ik]} S_{kj} + \Phi_{[jk]}S_{ki} \right) + C_{ikjl} S_{lk}
\ee 
It has recently been observed that this equation is actually valid for {\it any} solution admitting a multiple WAND, i.e., it is also valid for type II \cite{durkee}. Using the above expressions for $\Phi_{ij}$ and $S_{ij}$, this equation reduces to
\be
 \Phi \frac{\partial_r E}{E} \equiv 0.
\ee
If $\Phi \equiv 0$ then the traceless property of the Weyl tensor implies that $C_{ikjk}=0$. This implies that $C_{2\alpha 2\alpha} = C_{\alpha \gamma \beta \gamma} + C_{2\alpha 2 \beta} = 0$. But, by axisymmetry, $C_{2\alpha 2\beta} \propto \delta_{\alpha \beta}$ and $C_{\alpha \beta \gamma \delta} \propto (\delta_{\alpha \gamma} \delta_{\beta \delta} - \delta_{\beta \gamma} \delta_{\alpha \delta})$, so these equations imply that $C_{2\alpha 2\beta}=C_{\alpha  \beta \gamma \delta}=0$, i.e., $C_{ijkl}=0$. Therefore all Weyl components of boost weight zero vanish, implying that the spacetime is type III (or more special), a contradiction. Hence $\Phi \ne 0$, so we must have
\be
 \partial_r E \equiv 0.
\ee
We now substitute this result into the equation $R_{0\alpha 0\beta}=0$ following from the WAND condition (\ref{eqn:WANDcondition}). This gives $\partial_r B \left( \partial_z E - B \partial_v E \right) \equiv 0$. But $\partial_r B \ne 0$ (the WAND is non-geodesic) hence
\be
 \partial_z E - B \partial_v E \equiv 0.
\ee
Taking a $r$-derivative of this we obtain $\partial_v E = 0$. Plugging this back into the equation gives $\partial_z E = 0$. Hence $E$ is constant. Therefore the spacetime is a direct product $M_3 \times S^{d-3}$. This is only possible if $\Lambda>0$, then the Einstein equation implies that the solution must be $dS_3 \times S^{d-3}$ (which is type D).

In summary, we have shown that an axisymmetric multiple WAND of an axisymmetric type II or D vacuum spacetime must be geodesic unless the spacetime is $dS_3 \times S^{d-3}$. Combining this with the results for type III or N \cite{bianchi}, we learn that an axisymmetric multiple WAND of an axisymmetric vacuum spacetime must be geodesic unless the spacetime is $dS_3 \times S^{d-3}$ or type O.

\subsection{Solutions with a geodesic WAND}

The results Ref. \cite{bianchi} and the previous subsection establish that an axisymmetric multiple WAND in a vacuum solution is always geodesic (unless the spacetime is $dS_3 \times S^{d-3}$ or type O). In this subsection we shall determine all solutions with an axisymmetric geodesic WAND. We shall not assume that the WAND is a {\it multiple} WAND (so {\it a priori} the solution might be type I but we shall see that this does not happen).

We have $\partial_r B = 0$ because the WAND is geodesic. We can now introduce new coordinates $v'$ and $r'$ such that $v = v'+F(v',z)$, $r=r'G(v',z)$ for some functions $F,G$ that can be chosen to bring the metric to the same form as before but with $B \equiv 0$. Dropping the primes on the coordinates, we have
\be
 ds^2 = -U(v,r,z) dv^2 + 2 dv dr + 2 C(v,r,z) dv dz + D(v,r,z)^2 dz^2 + E(v,r,z)^2 d\Omega^2. \label{met:axigeo}
\ee
By rescaling, the WAND can be taken to be $\ell = \partial/\partial r$. We saw above that the null congruence associated with the WAND has vanishing rotation. Since it is geodesic, this implies that it is hypersurface orthogonal. In the above coordinates, it is orthogonal to hypersurfaces of constant $v$. Furthermore, $r$ is an affine parameter along the null geodesics.\footnote{If we assumed that $\ell$ is a {\it multiple} WAND then the $r$-dependence of the metric could be read off from Ref. \cite{nontwisting}. However we shall not make this assumption.}
There is some coordinate freedom remaining: the form of the metric is invariant under the transformations
\bea
\label{eqn:gauge}
 v &\rightarrow  & V(v), \qquad r \rightarrow r/\partial_v V, \nn 
 r &\rightarrow & r - F(v,z), \\
z &\rightarrow & z(v,z). \nonumber
\eea
All of this is well-known in the context of 4d solutions with a hypersurface orthogonal null geodesic congruence \cite{exactsolutionsbook}.

We shall employ the same null basis as before (i.e., (\ref{eqn:basis}) with $B \equiv 0$). The Riemann tensor of the above metric in this basis is given in Appendix \ref{app:riemann}. The WAND condition \eqref{eqn:WANDcondition} reduces to
\be
 \partial_r^2 D = \partial_r^2 E = 0,
\ee
Hence
\be
 D(v,r,z) = D_0(v,z) + rD_1(v,z), \qquad E(v,r,z) = E_0(v,z) + r E_1 (v,z),
\ee
for some functions $D_0,D_1,E_0,E_1$. The $00$ component of the Einstein equation is now automatically satisfied. Axisymmetry implies that the $0\alpha$ component is trivial. The $02$ component reduces to an equation linear in $C$:
\be
\label{eqn:errz}
 \partial_r^2 C - \left( \frac{\partial_r D}{D} - (d-3) \frac{\partial_r E}{E} \right) \partial_r C -
2(d-3) \frac{\partial_r D \partial_r E}{DE} C = 2(d-3)\left(\frac{\partial_r \partial_z E}{E} -\frac{\partial_r D \partial_z E}{DE}  \right).
\ee
This will determine the $r$-dependence of $C$. There are several different cases to consider.

\subsubsection{$E_1 \neq 0, \ D_1 \equiv0$}
We can use the residual freedom in $r$ and $z$ (equation \eqref{eqn:gauge}) to set $E_0 \equiv 0$ and $D_0 \equiv 1$, i.e. $D \equiv 1$. Then \eqref{eqn:errz} reduces to
\be
 \partial_r^2 C +  \frac{d-3}{r} \partial_r C= 2\frac{d-3}{r}\frac{\partial_z E_1}{ E_1},
\ee
which can be solved to give
\be
 C(v,r,z) = C_0(v,z) + \frac{C_1(v,z)}{(d-4)r^{d-4}} +\frac{2 \partial_z E_1}{E_1} r,
\ee
for arbitrary functions $C_0$ and $C_1$.  The $r$-dependence of $U(v,r,z)$ is determined by the $01$ component of Einstein's equation:
\bea
 U(v,r,z)&=& -\frac{C_1(v,z)^2}{2(d-4)^2 r^{2(d-4)}} - \frac{U_1(v,z)}{(d-4)r^{d-4}} + U_0(v,z)- \frac{2r}{E_1} \left( \partial_v E_1 + C_0 \partial_z E_1 \right) \nn &+& \frac{r^2}{d-2} \left( \frac{\partial_z^2 E_1}{E_1} -d \frac{(\partial_z E_1)^2}{E_1^2}- \Lambda \right) - \chi(r)  \left(\partial_z C_1 + \frac{(d^2-9d+22)}{d-4} C_1 \frac{\partial_z E_1}{E_1} \right)    ,
\eea
where $U_0$ and $U_1$ are arbitrary functions, and
\be
\chi(r)=\begin{cases} \log(r)\, , \hspace{16.9mm} d=5 \\ -\dfrac{1}{(d-5)r^{d-5}}\, , \quad d > 5 \end{cases}.
\ee
The $r$-dependence of the metric is now fully determined.
Comparing coefficients of terms with different $r$ dependence in the remaining components of the Einstein equation can be used to restrict the arbitrary functions above. The $\alpha \beta$ components of Einstein's equation give
\be
 C_1 = 0, \qquad \partial_z C_0 = 0,
\ee
\be
 U_0(v,z) = \frac{1}{E_1(v,z)^2} - C_0(v)^2,
\ee
The residual coordinate freedom $z \rightarrow z-f(v)$ can be used to set
\be
C_0 = 0.
\ee
The $22$ component of Einstein's equation gives
\be
\label{eqn:dgeoE1}
 \frac{\partial_z^2 E_1}{E_1} - 2 \frac{(\partial_z E_1)^2}{E_1^2} - \frac{\Lambda}{d-1} = 0.
\ee
Now the $12$ component of Einstein's equation implies
\be
\p_z U_1+2(d-3)\frac{\p_z E_1}{E_1} U_1=0,
\ee
which in turn implies
\be
U_1(v,z)=\frac{m(v)}{E_1(v,z)^{2(d-3)}},
\ee
for some arbitrary function $m(v)$.  The $11$ component of Einstein's equation then implies
\be
\p_vm(v)=(d-4) m(v) \frac{\p_vE_1(v,z)}{E_1(v,z)}. \label{eq:dgeom}
\ee
$m(v)=0$ implies that the spacetime is conformally flat (type O), so assume that $m(v) \neq 0$.  Then \eqref{eq:dgeom} implies
\be
 E_1(v,z) = \frac{|m(v)|^{1/(d-4)}}{g(z)},
\ee
for some positive function $g(z)$. Inserting this into equation \eqref{eqn:dgeoE1} gives the linear equation
\be
 g''(z) +\frac{\epsilon}{L^2} g(z)= 0,
\ee
where $L$ is defined by
\be
\label{eqn:Ldef}
\Lambda=\frac{(d-1) \epsilon}{L^2}, \qquad \epsilon \in \{-1,0,1\}.
\ee
Define a positive constant $\mu$ via the first integral
\be
 g'(z)^2 + \frac{\epsilon}{L^2} g(z)^2 = \eta \mu^2,
\ee
where $\eta \in \{-1,0,1\}$. 
Using the freedom to shift $z$ by a constant (and $z \rightarrow -z$) we have $g(z)=\mu G(z)$, where $G(z)$ is given by
\begin{center}
\begin{tabular}{l|lll}
$G(z)$ & $\eta=1$ & $\eta=0$ & $\eta=-1$ \\
\hline
$\epsilon=1$ & $L \sin (z/L)$ \\
$\epsilon=0$ &  $z$ & $\alpha$ \\
$\epsilon=-1$ &  $L \sinh (z/L)$ & $\alpha e^{\pm z/L}$ & $L \cosh (z/L)$
\end{tabular}
\end{center}
where $\alpha$ is a positive constant. Defining new coordinates $(V,\rho)$ by

\be
dV=\mu \frac{dv}{|m(v)|^{1/(d-4)}},  \qquad  \rho= \frac{r |m(v)|^{1/(d-4)}}{\mu G(z)^2},
\ee
the metric becomes 
\be
\label{eqn:string}
ds^2=dz^2+ G(z)^2\left[ -\left(1 - \frac{M}{\rho^{d-4}} -\eta \rho^2  \right) dV^2 +2dVd \rho +\rho^2 d \Omega^2 \right],
\ee
where $M = {\rm sign}(m(v)) \mu^{d-4}$. This is 
the warped product of a line, parametrized by $z$, with the $(d-1)$-dimensional Schwarzschild (anti)-de Sitter metric. The $\epsilon=\eta=0$ case is the Schwarzschild-Tangherlini black string. The $\epsilon=0$, $\eta=1$ case corresponds to taking a de Sitter slicing of Minkowski spacetime and replacing the de Sitter slices with the Schwarzschild-de Sitter metric. The $\epsilon=\eta=1$ case corresponds to doing the same thing for de Sitter spacetime. The $\epsilon=-1$ cases correspond to the same idea for slicings of anti-de Sitter space (In $d=5$, $\epsilon=-1$, $\eta=0$ is the AdS black string of Ref. \cite{chamblin}). A warped product whose Lorentzian factor is a type D Einstein space is also type D \cite{pravdatyped}. Hence these solutions are all type D.

\subsubsection{$E_1 \neq 0, \ D_1 \neq 0, \ D_0\equiv 0$: Robinson-Trautman solutions}

The coordinate freedom \eqref{eqn:gauge} can be used to set $E_0 \equiv 0$.
From (\ref{eqn:expmatrix}), these solutions have vanishing shear and non-vanishing expansion. Therefore they belong to the class of higher-dimensional Robinson-Trautman solutions \cite{ortaggio}. To give a self-contained presentation, we shall rederive these solutions here (with the additional restriction of axisymmetry). Using a transformation $z \rightarrow z'(z,v)$, we can set $D_1 \equiv 1$. The general solution to equation \eqref{eqn:errz} is
\be
 C(v,r,z) = C_1(v,z)r^2+\frac{C_2(v,z)}{r^{d-3}},
\ee
where $C_1$ and $C_2$ are arbitrary functions. But now $R_{22}-R_{\alpha\alpha}$ (no sum on $\alpha$) is independent of $U$ and, by the Einstein equation, must vanish. Equating coefficients of terms with different $r$-dependence gives $C_2 \equiv 0$ and
\be
 \frac{\partial_z^2 E_1}{E_1} - \frac{(\partial_z E_1)^2}{E_1^2} + \frac{1}{E_1^2}=0, \label{eqnarb:E1eq2}
\ee
\be
 \partial_z C_1 - \frac{\partial_z E_1}{E_1} C_1 + \frac{\partial_v E_1}{E_1} =0. \label{eqnarb:C1E1}
\ee
The $22$ component of the Einstein equation can then be solved to determine $U$:
\be
 U(v,r,z) = \frac{U_1(v,z)}{r^{d-3}} -\frac{\partial_z^2 E_1}{E_1} + 2r\partial_z C_1  - r^2 \left(C_1^2 + \frac{\Lambda}{d-1} \right),
\ee
where $U_1$ is an arbitrary function. Now examining the $12$ component of the Einstein equation gives $\partial_z U_1 =0$. The $11$ component of the Einstein equation reduces to
\be
 E_1 \partial_v U_1 + (d-1) \left(U_1 \partial_v E_1 - C_1 U_1 \partial_z E_1 \right)= 0. \label{eqnarb:E1U1}
\ee
If $U_1 \equiv 0$ then it can be shown that the above equations imply that the Weyl tensor vanishes hence the solution is type O. If $U_1 \ne 0$ then we can use the gauge freedom $v \rightarrow V(v)$ and $r \rightarrow r/\partial_v V$ to set $U_1 \equiv \mu$ for some non-zero constant $\mu$. Then (\ref{eqnarb:E1U1}) gives $\partial_v E_1 = C_1 \partial_z E_1$. From (\ref{eqnarb:C1E1}) we then learn that $\partial_z C_1 =0$. This implies that $C_1$ can be gauged away by a shift $z \rightarrow z - f(v)$. In the new gauge we have $\partial_v E_1=0$. The solutions of (\ref{eqnarb:E1eq2}) are $E_1 = R \sin (z/R)$, $z$ or $R \sinh (z/R)$ (using $z \rightarrow z-{\rm const}$ and $z \rightarrow -z$ to simplify) where $R$ is a positive constant. $R$ can be set to one by rescaling $z$, $v$ and $r$. The solution takes the final form
\be
\label{eqn:schw}
 ds^2 = -\left(k - \frac{M}{r^{d-3}} - \frac{\Lambda}{d-1} r^2 \right) dv^2 + 2 dv dr + r^2 d\Sigma_k^2,
\ee
where $M$ is a non-zero constant, and $d\Sigma_k^2$ is the metric on a $d-2$ dimensional space of unit constant curvature of sign $k$. This generalized Schwarzschild metric is of type D \cite{ortaggio}.

\subsubsection{$E_1 \neq 0, \ D_1 \neq 0, \ D_0 \neq 0$}

We use the transformations \eqref{eqn:gauge} to set $D_1 \equiv 1$ and $E_0 \equiv 0$. The general solution to equation \eqref{eqn:errz} is
\be
C(v,r,z) = (D_0(v,z)+r)^2 \left(C_1(v,z) + C_2(v,z) \int{\frac{dr}{r^{d-3}(D_0+r)^3}} \right)- \frac{D_0 \partial_z E_1}{E_1},
\ee
where $C_1$ and $C_2$ are arbitrary functions.
Now we consider the $22$ and $\alpha \alpha$ components of the Einstein equation. These equations are linear in $U$ and $\partial_r U$ and can be solved algebraically to determine $U$ and $\partial_r U$. The $r$-dependence is completely determined hence consistency of the solutions for $U$ and $\partial_r U$ gives an equation whose $r$ dependence is fully determined. Equating coefficients of terms with the same $r$ dependence then gives $C_2 \equiv 0$ together with
\bea
\partial_v D_0 &=& -\frac{1}{E_1^2}+\partial_z (D_0 C_1)+\frac{(\partial_z E_1)^2}{E_1^2} - \frac{\partial_z^2 E_1}{E_1}, \nn
\partial_v E_1 &=& \frac{\Lambda}{d-1} D_0 E_1 - E_1 \partial_z C_1 + C_1 \partial_z E_1.
\eea
The solution for $U$ is then
\bea
 U(v,r,z) &=& \frac{1}{E_1^2} - C_1^2 D_0^2 +2 C_1 D_0 \frac{ \partial_z E_1}{E_1} - \frac{(\partial_z E_1)^2}{E_1^2} \nn &+& \left(2 \partial_z C_1-\frac{2 \Lambda}{d-1} D_0 - 2 C_1^2 D_0  \right) r - \left( \frac{\Lambda}{d-1} + C_1^2 \right) r^2.
\eea
These results imply that the Weyl tensor vanishes. Hence these solutions are type O, i.e., Minkowski or (anti)-de Sitter spacetime.

\subsubsection{$E_1 \equiv 0$: Kundt solutions}

\label{sec:kundt}

Solving \eqref{eqn:errz} gives
\be
 C(v,r,z) = C_0(v,z) + \left( C_*(v,z) D_0(v,z) + 2(d-3)
 \frac{\partial_z E_0}{E_0} \right) r + \frac{1}{2} C_*(v,z) D_1(v,z)
 r^2,
\ee
where $C_0(v,z)$ and $C_*(v,z)$ are arbitrary functions. The $\alpha\alpha$ component of the Einstein equation does not involve $U$ so its $r$ dependence in completely determined. Equating coefficients of terms with different dependence on $r$ gives $D_1^3 [
(d-4)- \Lambda E_0^2]=0$. Hence either $D_1 \equiv 0$ or $E_0^2 \equiv (d-4)/\Lambda$. The
latter implies that spacetime is a direct product $M_3 \times S^{d-3}$, which requires $\Lambda>0$ and the spacetime must then be locally $dS_3 \times S^{d-3}$ which is of algebraic type D.

Assume instead that $D_1 \equiv 0$. We now
have $\partial_r D=\partial_r E=0$ so the geodesic congruence is free of expansion and
shear as well as twist. Spacetimes with vanishing expansion, shear and twist are referred to as Kundt spacetimes \cite{exactsolutionsbook,coley}. All vacuum Kundt solutions are type II or more special for any $d \ge 4$ \cite{ricci}. General $d$-dimensional Kundt spacetimes have been discussed recently \cite{kundt}. The general solution cannot be obtained in closed form. We shall now analyze such solutions assuming axisymmetry, which enables further progress to be made.

With $D_1 \equiv 0$, we can use the transformation $z \rightarrow z'(v,z)$ to set $D_0 \equiv 1$. Write the solution for $C$ as $C(v,r,z) = C_0(v,z) + r C_1(v,z)$. The shift $r \rightarrow r-F(v,z)$ has the effect $C_0 \rightarrow C_0 - \partial_z {F} - C_1 F$. Hence we can choose $F(v,z)$ to set $C_0 \equiv 0$. To summarize, we have brought the metric to the form
\be
 ds^2 = -U(v,r,z) dv^2 + 2 dv dr + 2 r C_1(v,z) dv dz + dz^2 + E_0(v,z)^2 d\Omega^2.
\ee
Some gauge freedom remains. The transfomations of the form \eqref{eqn:gauge} that preserve this form of the metric are 
\be
\label{eqn:zshift}
 v=v', \qquad z=z'+f(v'), \qquad r=r'+g(v',z'), \qquad \partial_{z'} g +C_1 g = -\partial_{v'} f,
\ee
\be
\label{eqn:vrescale}
 v=V(v'), \qquad r=\frac{r'}{\partial_{v'} V}, \qquad z=z'.
\ee  
The $\alpha\alpha$ component of the Einstein equation reduces to
\be
\label{eqnarb:kundtE0}
 \frac{\partial_z^2 E_0}{E_0} + (d-4) \frac{(\partial_z E_0)^2}{E_0^2} - C_1 \frac{\partial_z E_0}{E_0} -\frac{d-4}{E_0^2} + \Lambda = 0,
\ee
and the $22$ component of the Einstein equation reduces to
\be
\label{eqnarb:kundtC1}
 \partial_z C_1 - \frac{1}{2} C_1^2 - (d-3) \frac{\partial_z^2 E_0}{E_0} - \Lambda=0.
\ee
The $01$ component of the Einstein equation is satisfied if, and only if,
\be
 U(v,r,z) = U_0(v,z) + r U_1(v,z) + r^2 U_2(v,z),
\ee
where $U_0$ and $U_1$ are arbitrary functions, and
\be \label{kundt:U2}
 U_2(v,z) = \frac{1}{2} \partial_z C_1 + \frac{d-3}{2} C_1 \frac{\partial_z E_0}{E_0} - \frac{1}{2} C_1^2 - \Lambda.
\ee
The $12$ Einstein equation reduces to
\be \label{kundt:U1}
 \partial_z U_1 =- \partial_v C_1 - (d-3)C_1 \frac{\partial_v E_0}{E_0} - 2(d-3) \frac{\partial_v \partial_z E_0}{E_0}.
\ee
Finally, using the above equations, the $11$ Einstein equation reduces to
\be \label{kundt:U0}
 \partial_z^2 U_0  + \left( C_1 + (d-3) \frac{\partial_z E_0}{E_0} \right) \partial_z U_0 + \left( \partial_z C_1 + (d-3) C_1 \frac{\partial_z E_0}{E_0} \right) U_0 = 
(d-3)\left( 2\frac{\partial_v^2 E_0}{E_0} - U_1 \frac{\partial_v E_0}{E_0}\right).
\ee
As is familiar for Kundt solutions, the equations of motion separate into the ``background'' equations (\ref{eqnarb:kundtE0}) and (\ref{eqnarb:kundtC1}), which must be solved to determine $E_0$ and $C_1$. Given a solution of these equations, the other equations can be integrated to determine $U_0$ and $U_1$. The second step is trivial because the equations are linear. Hence solving the background equations is the non-trivial step that remains. However, the general solution to the background equations is not known analytically. 

Since the background equations do not involve $v$-derivatives, solving them equations is equivalent to solving the corresponding equations assuming that $E_0$ and $C_1$ are independent of $v$ and $U_0=U_1=0$. But in this case, the metric is static. In fact, we shall see below that the general type D axisymmetric Kundt metric is of this form. The background equations can only be solved in special cases e.g. the general solution with $d=4$ can be determined, and the general solution with $d=5$, $\Lambda=0$ and $U_2 = 0$ can also be obtained \cite{gregory}. Some time-dependent solutions based on the latter solution of the background equations were obtained in Refs \cite{jakimowicz1, jakimowicz2}.

It is convenient to define a positive function $W(v,z)$ by
\be
 W(v,z) = W_0(v) \exp (-\int^z C_1 (v,z') dz'),
\ee
where $W_0(v)$ is an arbitrary positive function, so
\be
 C_1 = - \frac{\partial_z W}{W}. \label{kundt:W}
\ee
The background equations become
\be
\frac{\partial_z^2 E_0}{E_0} + (d-4) \frac{(\partial_z E_0)^2}{E_0^2} +\frac{\p_zW}{W} \frac{\partial_z E_0}{E_0} -\frac{d-4}{E_0^2} + \Lambda = 0,
\ee
\be
\frac{1}{2}\frac{(\p_zW)^2}{W^2} -\frac{\p^2_zW}{W} - (d-3) \frac{\partial_z^2 E_0}{E_0} - \Lambda=0.
\ee
Equation \eqref{kundt:U0} becomes
\be
 U_2 = -\frac{\p_z^2 W}{2 W} - \frac{(d-3)\p_z W \p_z E_0}{2 WE_0} - \Lambda.
\ee
These equations imply that
\be
 \partial_z \left( W U_2 \right) = 0,  \label{U2C1eqn}
\ee
hence we can always choose $W_0(v)$ so that 
\be
 U_2(v,z) = -\frac{k}{W(v,z)}, \label{kundt:U2W}
\ee
with $k \in \{1,0,-1\}$. We can now define a new coordinate $R$ by
\be
 r = W(v,z) R.  \label{kundt:coordR}
\ee
The metric becomes
\be
\label{eqn:newkundt}
 ds^2 = W(v,z) \left\{ - \left[ \frac{U_0(v,z)}{W(v,z)}+ R \left(U_1 (v,z)-2\frac{\partial_v W}{W} \right) -k  R^2 \right] dv^2 + 2 dv dR \right\} + dz^2 + E_0(v,z)^2 d\Omega^2,
\ee
We now consider a further classification of the Kundt solutions according to their algebraic type. Using the above equations to simplify the Weyl tensor, we find that the only independent nonzero components are:
\be
C_{0 \alpha 1 \beta}= \delta_{\alpha \beta} \left(C_1 \frac{\p_z E_0}{2E_0} -\frac{\Lambda}{d-1}\right),
\ee
\be
C_{\alpha \beta \gamma \delta}=2\delta_{\alpha [\delta} \delta_{\gamma] \beta}\left(\frac{\Lambda}{d-1} +\frac{(\p_zE_0)^2}{E_0^2}-\frac{1}{E_0^2} \right),
\ee
\be \label{kundt:C1ttd}
C_{1 \alpha \beta 2}=\frac{\delta_{\alpha \beta}}{2}\left( C_1 \frac{\p_v E_0}{E_0}+2\frac{\p_v \p_zE_0}{E_0}\right)=-\frac{\delta_{\alpha \beta}}{2(d-3)}(\p_zU_1+\p_vC_1),
\ee 
\be
C_{1 \alpha 1 \beta}=\delta_{\alpha \beta} \left\{-\frac{1}{2(d-3)} \p_z (\p_z U_0+C_1 U_0)
 +\left[U_2\frac{\p_vE_0}{E_0}+\left(\frac{1}{2} \p_z U_1+\p_vC_1 \right)\frac{\p_zE_0}{E_0} \right]r\right\}. \label{kundt:C1t1t}
\ee
Note that while $C_{0101}, \, C_{0112}, \, C_{0212}, \, C_{1212}$ and $C_{\alpha 2 \beta 2}$ are nonzero, they are related to the above components by the tracefree property of the Weyl tensor.  The first two Weyl components written above are of boost weight 0, while the remaining two are of boost weight -1 and -2 respectively. Hence the solutions we are considering here are at least type II, confirming the general result of Ref. \cite{ricci}.

\medskip

\noindent {\bf Type III and N}

\medskip

Consider the case in which the solution is type III, or more special. In this case, the Weyl components of boost weight 0 vanish. This gives, for $d>4$, the following equations
\be
C_1 \frac{\p_z E_0}{E_0}=\frac{2\Lambda}{d-1}, \label{KundtIII:eqn1}
\ee
\be
\frac{(\p_zE_0)^2}{E_0^2}=\frac{1}{E_0^2}-\frac{\Lambda}{(d-1)}. \label{KundtIII:eqn2}
\ee
Note that equation \eqref{KundtIII:eqn2} is not present for $d=4$. Solving this equation gives
\be
\label{eqn:typeNE0}
E_0(v,z)=\begin{cases} L \sin (z/L) & \text{if $\Lambda>0$,}   \\ z & \text{if $\Lambda=0$,}  \\ L \sinh (z/L) & \text{if $\Lambda<0$,}  \end{cases}
\ee
where $L>0$ is defined by \eqref{eqn:Ldef}, and we have used the freedom \eqref{eqn:zshift} to eliminate an arbitrary function of $v$ (we've also fixed signs using $z \rightarrow \pm z$). Equation \eqref{KundtIII:eqn1} now determines $C_1$:
\be
\label{eqn:typeNC1}
C_1(v,z)=\begin{cases} (2/L) \tan(z/L) & \text{if $\Lambda>0$,} \\ 0 & \text{if $\Lambda=0$,} \\ -(2/L) \tanh (z/L) & \text{if $\Lambda<0$.}  \end{cases}
\ee
Since $E_0$ is independent of $v$, equation \eqref{kundt:C1ttd} gives that $C_{1 \alpha \beta 2}=0$, hence these solutions are type N or O. There are no axisymmetric type III Kundt solutions for $d>4$. However, such solutions do exist for $d=4$ \cite{exactsolutionsbook}.

Continuing the analysis, note that the coefficient of $r$ in $C_{1 \alpha 1 \beta}$ (given by \eqref{kundt:C1t1t}) vanishes, and so $C_{1 \alpha 1 \beta}$ reduces to
\be
\label{eqn:kundttypeOcond}
C_{1 \alpha 1 \beta}=-\frac{\delta_{\alpha \beta}}{2(d-3)} \p_z (\p_z U_0+C_1 U_0).
\ee
$U_2(v,z)$ can be calculated using \eqref{kundt:U2}:
\be
U_2(v,z)= \begin{cases} -(1/L^2) \sec^2 (z/L) & \text{if $\Lambda>0$,} \\ 0 & \text{if $\Lambda=0$,} \\ (1/L^2) \, \text{sech}^2 (z/L) & \text{if $\Lambda<0$.}
      \end{cases}
\ee
Equation \eqref{kundt:U1} implies that $U_1(v,z)$ is independent of $z$: $U_1(v,z)=U_1(v)$. We can then use the transformation $v \rightarrow V(v)$, $r \rightarrow r/\partial_v V$ to arrange that
\be
 U_1 \equiv 0.
\ee
Finally, equation \eqref{kundt:U0} can be solved to determine $U_0$. For $\Lambda=0$, the solution is, for $d>4$
\be
 U_0 = \frac{F(v)}{z^{d-4}} + G(v),
\ee
(for $d=4$, the first term is replaced by $F(v) \log z$). A shift $r \rightarrow r - f(v)$ can be used to set $G(v) \equiv 0$. The only independent non-zero component of the Weyl tensor is (\ref{eqn:kundttypeOcond}). This reveals that the solution is type O if, and only if, $F(v) \equiv 0$. Therefore, the general axisymmetric type N Kundt solution with $\Lambda=0$, is given by the following metric (for $d>4$)
\be
\label{eqn:ppwave}
ds^2=-\frac{F(v)}{z^{d-4}}dv^2+2dvdr+dz^2+z^2d\Omega^2.
\ee
The null vector field $\ell$ is covariantly constant, and so the solution above belongs to the family of pp-waves \cite{ehlers}.

For $\Lambda < 0$, the solution for $U_0$ is
\be
 U_0 = \cosh^2 (z/L) \left[ F(v) I_-(z) + G(v) \right],
\ee
where
\be
 I_-(z) = \int_z^\infty \frac{dz}{\cosh^2 (z/L) \sinh^{(d-3)} (z/L)},
\ee
Define a new coordinate $R$ by
\be
 r = R \cosh^2 (z/L).
\ee
The metric becomes
\be
 ds^2 = \cosh^2 (z/L) \left[ - (F(v) I_-(z) + G(v) + R^2/L^2 ) dv^2 + 2 dv dR \right] + dz^2 + L^2 \sinh^2 (z/L) d\Omega^2.
\ee
Now the transformations $R \rightarrow R - f(v)$ followed by $v \rightarrow V(v)$, $R \rightarrow R/\partial_v V$ can be used to eliminate $G(v)$, giving the final form of the solution:
\be
\label{eqn:adswave}
 ds^2 = \cosh^2 (z/L) \left[ - (F(v) I_-(z) + R^2/L^2 ) dv^2 + 2 dv dR \right] + dz^2 + L^2 \sinh^2 (z/L) d\Omega^2.
\ee
A similar analysis for $\Lambda >0$ (or $L \rightarrow i L$) gives
\be
\label{eqn:dswave}
 ds^2 = \cos^2 (z/L) \left[ - (F(v) I_+(z) - R^2/L^2 ) dv^2 + 2 dv dR \right] + dz^2 + L^2 \sin^2 (z/L) d\Omega^2,
\ee
where
\be
 I_+(z) = \int \frac{dz}{\cos^2 (z/L) \sin^{d-3} (z/L)}.
\ee
These solutions are type O, i.e., isometric to (anti-)de Sitter space, if, and only if, $F(v) \equiv 0$. If $F(v)$ is not identically zero then these metrics are the general axisymmetric type N Kundt solutions for $\Lambda \ne 0$, $d>4$.\footnote{For $d=4$, these solutions are a special case of more general type N Kundt solutions discussed in Ref. \cite{typeN}.}  It seems natural to interpret the type N solutions as describing gravitational waves propagating in a type O background.

\medskip

\noindent {\bf Type D}

\medskip

Now consider type D solutions, for which there exists a second multiple WAND $n'$. Note that $n'$ need not coincide with the null basis vector $n$ defined above. If $n'$ were not axisymmetric then the solution would be encompassed by the analysis of section \ref{sec:WANDnotaxi}. However, the results of that section reveal that, in this case, {\it both} multiple WANDs would fail to be axisymmetric, which is not the case here. Hence we can assume that $n'$ is axisymmetric. The most general form it can take is
\be
 n'=n -\frac{1}{2}a(v,r,z)^2  \, \ell+a(v,r,z) \, m_2
\ee
where $a(v,r,z)$ is arbitrary. Let us change to a new null frame $(\ell',n',m_i')$, with 
\be
  \ell'=\ell, \qquad m_2' = m_2 - a \ell, \qquad m_\alpha'=m_\alpha.
\ee
Note that $a\equiv 0$ corresponds to the frame used above. The fact that $\ell$ is a multiple WAND guarantees that Weyl components of boost weight 0 are the same in the two frames. The negative boost weight components in the new frame are related to the components in the old frame by
\bea
&C'_{1\alpha \beta 2}=C_{1\alpha \beta 2}-aC_{2\alpha 2\beta }+aC_{0\alpha 1\beta },\notag \\ \label{Weylnframe}
&C'_{1\alpha 1 \beta}=C_{1\alpha 1 \beta}-2aC_{1\alpha \beta 2}-a^2C_{0\alpha 1 \beta}+a^2C_{2\alpha 2 \beta}.
\eea
Type D solutions are those for which Weyl tensor components of boost weight $-2$ and $-1$ (in the new frame) vanish, giving
\be
C_1\p_vE_0+2\p_v\p_zE_0+a\left(C_1\p_zE_0+2\p^2_zE_0 \right)=0, \label{weyl-1eqn}
\ee
\be
C_{1\alpha 1 \beta}-2aC_{1\alpha \beta 2}-a^2C_{0\alpha 1 \beta}+a^2C_{2\alpha 2 \beta}=0,  \label{weyl-2eqn}
\ee
where the second equation has not been written explicitly for brevity.  Note that equation \eqref{weyl-1eqn} implies either $a=a(v,z)$; or $C_{2\alpha 2 \beta} = C_{0 \alpha 1 \beta}$ and $C_{1\alpha \beta 2} = 0$. In the latter case, equation \eqref{weyl-2eqn} implies $C_{1\alpha 1 \beta}=0$, and then one finds that $n$ is a multiple WAND, i.e., one can set $a \equiv 0$. Hence, in either case, we have $a=a(v,z)$. Therefore, in equation \eqref{weyl-2eqn}, the only term with $r$ dependence is that contained in $C_{1\alpha 1 \beta}$, which must vanish, giving (from equation \eqref{kundt:C1t1t})
\be
2U_2\frac{\p_vE_0}{E_0}+\left( \p_z U_1+2\p_vC_1 \right)\frac{\p_zE_0}{E_0}=0. \label{dcon1}
\ee
To simplify the analysis,  assume that the spacetime is not $dS_3 \times S^{d-3}$ (which we already know is type D). The results of section \ref{Dimpliesgeo} imply that the second multiple WAND $n'$ must be geodesic. Axisymmetry implies that the geodesic equation reduces to 
\be
 {m_2'}^a {n'}^b \nabla_b n'_a=0.
\ee
The LHS is linear in $r$ so this gives two equations:
\be
2a\p_za+2\p_va+a^2C_1-aU_1+\p_zU_0+C_1U_0=0, \label{ngeo1}
\ee
\be
\p_zU_1+2\p_vC_1-2aU_2=0. \label{ngeo2}
\ee
Proceed by simplifying equation \eqref{dcon1} using equation \eqref{ngeo2}:
\be
2U_2\left(a+ \frac{\p_vE_0}{\p_zE_0}\right) =0. \label{typed:a}
\ee
Note that $\p_z E_0$ is not identically zero, since otherwise eq. \eqref{eqnarb:kundtE0} implies that $E_0$ is constant, which gives $dS_3 \times S^{d-3}$. There are two cases.

{\it Case 1.} $\p_v E_0=-a \p_z E_0.$
Using this to eliminate $\p_v E_0$ from eq.  \eqref{weyl-1eqn} gives
$\partial_z a=0$,
 so $a=a(v)$. Now we can use a transformation of the form \eqref{eqn:zshift} to reach a gauge in which $\partial_v E_0 \equiv 0$, i.e., $a \equiv 0$. Substituting \eqref{eqnarb:kundtC1} into \eqref{ngeo2} then gives $\partial_v C_1 = 0$. Now equation \eqref{kundt:U1} gives $\partial_z U_1 = 0$. Define a positive function $W(z)$ by equation \eqref{kundt:W}. Equation \eqref{kundt:U2} reveals that $U_2$ is independent of $v$ so equation \eqref{U2C1eqn} implies \eqref{kundt:U2W} as before (using the freedom to rescale $W$ by  a constant). Equation \eqref{ngeo1}  gives $U_0 = F(v) W(z)$. Defining the coordinate $R$ by \eqref{kundt:coordR}, the metric can be brought to the form \eqref{eqn:newkundt}:
 \be
 ds^2 = W(z) \left\{ - \left[ F(v)+ R U_1 (v)  -k  R^2 \right] dv^2 + 2 dv dR \right\} + dz^2 + E_0(z)^2 d\Omega^2.
\ee
The transformation $R \rightarrow R - G(v)$ can be used to set $F(v) \equiv 0$, then a transformation $v \rightarrow V(v)$, $R \rightarrow R/\p_v V$ can be used to set $U_1 \equiv 0$. The metric is then
\be
\label{kundtD}
 ds^2 = W(z) d\Sigma^2 + dz^2 + E_0(z)^2 d\Omega^2,
\ee
where $d\Sigma^2$ is the metric  on a 2d Lorentzian space with Ricci scalar $2k$, i.e., Minkowski or (anti-) de Sitter spacetime.

{\it Case 2.} $U_2 \equiv 0$. Equations \eqref{kundt:U2} and \eqref{eqnarb:kundtC1} imply
\be
\p_z \left( \frac{C_1}{E_0^{d-3}} \right) = 2(d-3) \frac{\partial_z^2 E_0}{E_0^{d-2}}, \label{U20z}
\ee
while equations \eqref{ngeo2} and \eqref{kundt:U1} imply
\be
\p_v \left( \frac{C_1}{E_0^{d-3}} \right) = 2(d-3) \frac{\partial_v \p_z E_0}{E_0^{d-2}} \label{U20v}
\ee
The integrability condition for these equations is
\be 
 \p_z \left( \frac{\p_v E_0}{\p_z E_0} \right) = 0,
\ee
hence $\p_v E_0 = h(v) \p_z E_0$ for some function $h(v)$. A gauge transformation of the form \eqref{eqn:zshift} can be used to reach a gauge in which $\p_v E_0 \equiv 0$, i.e., $h(v) \equiv 0$. Equation \eqref{U20v} now gives $\p_v C_1 \equiv 0$ and \eqref{kundt:U1} gives $\p_z U_1 \equiv 0$. Now if $a \equiv 0$ then we are back to case 1, so assume $a \ne 0$. Then the coefficient of $a$ in equation \eqref{weyl-1eqn} must vanish. But this is the case discussed below equation \eqref{weyl-2eqn}, where $n$ is a multiple WAND, so one can set $a \equiv 0$ after all, leading back to case 1.

In summary, we have shown that, for a general type D axisymmetric Kundt metric, one can find a gauge in which $E_0$ and $C_1$ are independent of $v$ and $U_0 = U_1 = 0$. The metric can be transformed to the form \eqref{kundtD}.\footnote{
Note that the special case $dS_3 \times S^{d-3}$ can be written in the form \eqref{kundtD} (with constant $E_0$ and $k=1$).}
 Conversely, the warped product structure of \eqref{kundtD} implies that any such solution is type D or O \cite{pravdatyped}. Type O corresponds to the solutions for $E_0(z)$ and $C_1(z)$ found in our discussion of type N, i.e., equations \eqref{eqn:typeNE0}, \eqref{eqn:typeNC1}. 

\section{Axisymmetric solutions with a non-axisymmetric WAND} 

\label{sec:WANDnotaxi}

Consider the Kaluza-Klein bubble spacetime \cite{witten} (generalized to include $\Lambda$) obtained by analytic continuation of the Schwarzschild solution:
\be
\label{eqn:bubble}
 ds^2 = r^2 ds^2 (dS_{d-2}) + \frac{dr^2}{U(r)} + U(r) dz^2, \qquad U(r) = 1 - \frac{m}{r^{d-3}}-\frac{\Lambda r^2}{d-1},
\ee
where $m \ne 0$ is a constant, and $dS_{d-2}$ is $(d-2)$ dimensional de Sitter space:
\be
ds^2(dS_{d-2}) = -dt^2 + \cosh^2 t \, d\Omega^2.
\ee
This spacetime is obviously axisymmetric. It is a warped product of $dS_{d-2}$ and $R^2$ and is hence type D \cite{pravdatyped}. We did not discover this spacetime above. This is because the multiple WANDs live in the $dS_{d-2}$ directions and hence must have non-vanishing components along $S^{d-3}$, i.e., they are not axisymmetric. Here this is possible because the axisymmetry $SO(d-2)$ is part of the bigger $SO(1,d-2)$ de Sitter symmetry. We shall show that this symmetry enhancement is necessary for a {\it multiple} WAND to be non-axisymmetric.

Consider first the case in which we have a non-axisymmetric (multiple) WAND $\ell$ that is everywhere orthogonal to the $S^{d-3}$ orbits of $SO(d-2)$, i.e., the only non-zero components of the WAND (in the coordinates of \eqref{eqn:orthogonal}) are $\ell^a=\ell^a(x,\Omega)$, where $\Omega$ refers to the coordinates on $S^{d-3}$. Now, since the Weyl tensor is axisymmetric, it is clear that $\ell^a(x,\Omega_0)$ is also a (multiple) WAND where $\Omega_0$ is an arbitrary point on $S^{d-3}$. But this new WAND does not vary on $S^{d-3}$, i.e., it is axisymmetric. Hence we conclude that, if the WAND is everywhere orthogonal to $S^{d-3}$ then there is no loss of generality in assuming that it is axisymmetric. 

Assume instead that we have an axisymmetric spacetime with metric \eqref{eqn:orthogonal} and that a WAND $\ell$ is {\it not} orthogonal to $S^{d-3}$ at some point. Then the same must hold in a neighbourhood of that point. Consider the ``unphysical'' spacetime $M_3 \times S^{d-3}$ with the product metric obtained by multiping \eqref{eqn:orthogonal} by $E(x)^{-2}$. We shall work with this spacetime for most of this section. Obviously $\ell$ is a WAND of this spacetime. By rescaling $\ell$ we can ensure that the projection of $\ell$ onto $S^{d-3}$ is a unit vector (in our neighbourhood). Hence we can write $\ell=(e_0+e_3)/\sqrt{2}$ where $e_0$ is a timelike unit vector in $M_3$ and $e_3$ a unit vector on $S^{d-3}$. Choose $n=(-e_0+e_3)/\sqrt{2}$. Choose $e_1$ and $e_2$ so that $\{e_0,e_1,e_2\}$ is an orthonormal basis for $M_3$, and choose $e_4 \ldots e_{d-1}$ so that $\{e_3 \ldots, e_{d-1}\}$ is an orthonormal basis for $S^{d-3}$. Now take $\{m_i\} = \{e_1,e_2,e_4,\ldots, e_{d-1} \}$. Let $\hat{a}, \hat{b}$ take values $1,2$ and let $\hat{\alpha}, \hat{\beta}$ take values $4 \ldots, (d-1)$ and $\alpha,\beta$ take values $3, \ldots, d-1$. By axisymmetry we have that $C_{abcd}$ vanishes if there are an odd number of indices of the form $\alpha,\beta$. We also have
\be
 C_{0\alpha 0 \beta} = a \delta_{\alpha \beta}, \qquad C_{\hat{a} \alpha \hat{b} \beta} = C_{\hat{a} \hat{b}} \delta_{\alpha\beta}, \qquad C_{\alpha \beta \gamma \delta} =b \left( \delta_{\alpha \gamma} \delta_{\beta \delta} - \delta_{\alpha \delta} \delta_{\beta\gamma} \right),
\ee
for some quantities $a$, $C_{\hat{a} \hat{b}}$, $b$. Now the WAND condition $C_{\mu\nu\rho\sigma} \ell^\mu m_i^\nu \ell^\rho m_j^\sigma=0$ reduces to
\be
 C_{\hat{a} \hat{b}} = - C_{0 \hat{a} 0 \hat{b}}, \qquad b=-a.
\ee
The first equation follows from choosing $i,j=\hat{a},\hat{b}$ in the WAND condition and the second by choosing $i,j,=\hat{\alpha},\hat{\beta}$.

Now, since we have a product metric, $C_{abcd}$ is fully determined by the Ricci tensor of $M_3$. Hence these conditions give conditions on this Ricci tensor. Using the formulae in \cite{pravdatyped} (and $R_{\alpha\beta}=(d-4)\delta_{\alpha\beta}$), we find that the Ricci tensor of $M_3$ must obey
\be
 R_{00} = 2, \qquad R_{\hat{a} \hat{b}} = \frac{1}{2} R_{\hat{c} \hat{c}} \, \delta_{\hat{a} \hat{b}}.
\ee
Similarly, the additional condition for $\ell$ to be a multiple WAND reduces to
\be
 R_{0\hat{a}}=0.
\ee
Note that these conditions are invariant under $e_0 \rightarrow - e_0$, which implies that if $\ell$ is a multiple WAND then so is $n$. Hence the spacetime is type D or more special. From now on, we assume that $\ell$ is indeed a multiple WAND.  Note that we can argue as we did in the second paragraph of this section to deduce that there is no loss of generality in assuming that $e_0$ is axisymmetric, which we shall assume henceforth.

Using capital letters $M,N, \ldots$ to denote coordinate indices in $M_3$, we can summarize the form of the 3d Ricci tensor as
\be
\label{eqn:3dRicci}
 R_{MN} = (\mu+1) u_M u_N + (\mu-1) g_{MN},
\ee
where $u^M \equiv e_0^M$. This is the Ricci tensor that would arise from a solution of the 3d Einstein equations sourced by a perfect fluid with energy density $\mu$ and pressure $p\equiv 1$ (with $8\pi G_3 = 1$). The contracted Bianchi identity (or stress tensor conservation) gives
\be
 (\mu+1) u \cdot \nabla u^M=0,
\ee
and
\be
\label{eqn:dTmu}
 (\mu+1) \theta + u \cdot \nabla \mu=0,
\ee
where the expansion $\theta$ is defined by $\theta = \nabla \cdot u$.
The first of these equations implies that either $\mu \equiv -1$ or $u^M$ is tangent to affinely parametrized geodesics in $M_3$. In the former case, we have $R_{MN} = -2g_{MN}$, which implies that $M_3$ is locally isometric to $AdS_3$ with unit radius, which implies that $M_3 \times S^{d-3}$ is conformally flat, so the physical spacetime is type O. Assume henceforth that this is not the case, so $\mu \ne -1$ and $u^M$ is geodesic in $M_3$.

The Einstein equation for the physical metric $E^2 g$ is
\be
\Lambda \delta^\mu_\nu = E^{-2} R^\mu_\nu + (d-2) E^{-1} \nabla^\mu \nabla_\nu E^{-1}  - \frac{1}{d-2} E^{-d} \nabla^2 E^{d-2} \delta^\mu_\nu,
\ee
where $\nabla$ is the covariant derivative with respect to the unphysical metric $g$, and indices are raised and lowered with this metric. The components tangent to the sphere give
\be
 \frac{1}{d-2} E^{-d} \nabla^2 E^{d-2} = (d-4) E^{-2} - \Lambda.
\ee
Using this, and \eqref{eqn:3dRicci}, the components tangent to $M_3$ give
\be
\label{eqn:omegaeq}
 E^{-1} \left[ (\mu+1) u_M u_N + (\mu - (d-3)) g_{MN} \right] + (d-2) \nabla_M \nabla_N E^{-1} = 0.
\ee
We now act on this with $\nabla_P$, antisymmetrize on $MP$, and use the fact that the Riemann tensor in 3d is determined by the Ricci tensor, which is given by \eqref{eqn:3dRicci}. This results in the equation
\bea
 \label{eqn:bigeq}
 0 &=& (\mu+1) \nabla_{[P} u_{M]} u_N -(d-3)(\mu+1) \nabla_{[P} E^{-1} g_{M]N} + E^{-1} (\nabla_{[P} \mu) u_{M]} u_N + E^{-1} (\mu+1) B_{[MP]} u_N \nonumber \\
 &+& E^{-1} (\mu+1) B_{N[P} u_{M]} + E^{-1} (\nabla_{[P} \mu) g_{M]N} - (d-2) (\mu+1) g_{N[M} u_{P]} u \cdot \nabla E^{-1} \nonumber \\ &-& (d-2) (\mu+1) \nabla_{[P} E^{-1} u_{M]} u_N
\eea
where
\be
 B_{MN} = \nabla_N u_M.
\ee
Contracting with $u^N$, this equation reduces to $(\mu+1)B_{[MP]}= 0$. Since we are assuming that $\mu +1$ is not identically zero, we must have
\be
 B_{[MN]} = 0,
\ee
i.e. $du=0$, so $u^M$ is hypersurface-orthogonal. Define the projector 
\be
h_{MN} = g_{MN} + u_M u_N
\ee
and now contract \eqref{eqn:bigeq} with $h^N_Q$ to get
\bea
\label{eqn:bigeq2}
 0 &=& -(d-3) (\mu+1) \nabla_{[P} E^{-1} h_{M]Q} + E^{-1} (\nabla_{[P} \mu) h_{M]Q} + E^{-1} (\mu+1) \hat{B}_{Q[P} u_{M]}\nonumber \\ &-& (d-2) (\mu+1) h_{Q[M} u_{P]} u \cdot \nabla E^{-1},
\eea
where
\be
 \hat{B}_{MN} = h_M^P h_N^Q B_{PQ}.
\ee
We can define the expansion and shear of the geodesic congruence tangent to $u^M$ in terms of the trace and traceless parts of $\hat{B}_{MN}$:
\be
 \hat{B}_{MN} = \frac{1}{2} \theta h_{MN} + \sigma_{MN}.
\ee
Contracting \eqref{eqn:bigeq2} with $u^M$ gives $(\mu+1) \hat{B}_{QP} \propto h_{PQ}$,
hence the congruence is shear-free:
\be
 \sigma_{MN} = 0.
\ee
Equation \eqref{eqn:bigeq2} now reduces to
\be
\label{eqn:Xeq}
 X_{[P} h_{M]Q} = 0,
\ee
where
\be
 X_P = - (d-3) (\mu+1) \nabla_P E^{-1} + E^{-1} \nabla_P \mu - \frac{1}{2} \theta E^{-1} (\mu+1) u_P - (d-2) (\mu+1) u_P u \cdot \nabla E^{-1}.
\ee
However, contracting \eqref{eqn:Xeq} with $h^P_N$ reveals that $X_P = 0$. Decomposing this into a part orthogonal to $u^M$ and a part parallel to $u^M$ gives
\be
\label{eqn:orthog} 
 h_M^N \nabla_N \left( E^{d-3} (\mu+1) \right)=0,
\ee
\be
\label{eqn:thetaomega}
 \theta = -2 E^{-1} u \cdot \nabla E
\ee
where \eqref{eqn:dTmu} was used to simplify the second equation.

Let $\Sigma_0$ be a surface orthogonal to $u^M$ (recall that $u^M$ is hypersurface orthogonal), let $x^i$ be coordinates on $\Sigma_0$. Assign coordinates $(T,x^i)$ to the point proper time $T$ along the geodesic tangent to $u^M$ starting at the point on $\Sigma_0$ with coordinates $x^i$. In this chart, the metric is
\be
 ds^2 = -dT^2 + h_{ij}(T,x) dx^i dx^j,
\ee
and $u=\partial/\partial T$. From the definition of $\hat{B}_{MN}$ and using the fact that the rotation and shear of the geodesics vanish, and equation \eqref{eqn:thetaomega} we deduce that
\be
 h_{ij}(T,x) = E^{-2} H_{ij}(x),
\ee
for some 2-metric $H_{ij}$ independent of $T$. Eliminating $\theta$ between equations \eqref{eqn:dTmu} and \eqref{eqn:thetaomega} gives $(\mu+1) = f(x) E^2$ for some (non-zero) function $f$. Substituting this into equation \eqref{eqn:orthog} and integrating gives
\be
\label{eqn:omegafg}
 E^{d-1} = \frac{g(T)}{f(x)},
\ee
for some function $g(T)$. Now contracting \eqref{eqn:omegaeq} with $u^M u^N$ gives
\be
 \partial_T^2 E^{-1} + E^{-1} = 0.
\ee
Using \eqref{eqn:omegafg} and the freedom to shift $T$ by a constant we can solve to obtain
\be
 E^{-1} = r(x)^{-1} \sin T,
\ee
for some non-zero function $r(x)$.

 Putting everything together, the physical metric is
\be
 ds^2 = r(x)^2 \left( \frac{-dT^2 + d\Omega^2}{\sin^2 T} \right) + H_{ij}(x) dx^i dx^j.
 \ee
The metric in brackets is the metric of $(d-2)$-dimensional de Sitter space. The full metric is invariant under the de Sitter isometry group. Hence if we Wick rotate to Euclidean signature then obtain a spherically symmetric spacetime so we can apply Birkhoff's theorem to deduce that the above metric must be either the Kaluza-Klein bubble spacetime \eqref{eqn:bubble}, or (if $r(x)$ is constant and $\Lambda>0$)  $dS_{d-2} \times S^2$.

\bigskip

\begin{center} {\bf Acknowledgments} \end{center}

\noindent

We thank J. Santos for discussions and V. Pravda for comments on a draft manuscript. We are especially grateful to M. Durkee for discussions and comments on a draft manuscript. MG is supported by EPSRC. HSR is a Royal Society University Research Fellow.

\appendix

\section{Curvature tensors for axisymmetric metrics with axisymmetric geodesic WAND}

\label{app:riemann}

In this Appendix, we record the non-zero components of the Riemann tensor of the metric \eqref{met:axigeo} describing an axisymmetric spacetime with an axisymmetric geodesic WAND, using the null basis \eqref{eqn:basis} (with $B \equiv 0$)
\be
R_{0101}= \frac{1}{4}\left(\frac{(\p_r C)^2}{D^2}+2\p_r^2 U\right),
\ee

\be
R_{0102}= -\frac{1}{2D}\left(\p_r^2 C - \frac{\p_r D}{D} \p_r C \right) ,
\ee

\be
R_{0202}= -\frac{\p_r^2 D}{D},
\ee

\be
R_{0 \alpha 0 \beta}= - \delta_{\alpha \beta} \frac{\p_r^2 E}{E},
\ee

\begin{align}
R_{1012}=-\frac{1}{4D^2} &\left[ 2CD \p^2_{r}U- DU\p^2_rC -2D\p_r\p_zU -2D\p_{v}\p_{r}C -2(C \p_rU-\p_zU)\p_rD \right.\notag \\ & \left. +4\p_vC\p_rD -2\p_rC\p_vD +U\p_rC\p_rD \right], 
\end{align}

\begin{align}
R_{1212}=\frac{1}{4D^3} &\left[2D(C^2\p_r^2U+ \p_z^2U)- 2CDU\p_r^2C -D^2(4 \p_v^2D+ U^2\p_r^2D)- 4CD\p_{r}\p_{z}U - 4D^2U \p_{v}\p_{r} D\right. \notag \\
& -2D(2C\p_{v}\p_{r}C-2\p_{v}\p_{z}C-U\p_{r}\p_{z}C)- 2C^2 \p_rU\p_rD + 2CU\p_rC \p_rD -2\p_zU\p_zD  \notag \\
&-2D^2(\p_vU\p_rD -\p_rU\p_vD) -2D(\p_rU\p_zC -\p_zU\p_rC) +2C(\p_rU\p_zD +\p_zU \p_rD) \notag \\
&\left. + 4 \p_vC (C\p_rD-\p_zD)- 2U \p_rC\p_zD  \right],
\end{align}

\begin{align}
R_{1 \alpha 1 \beta}=  \frac{\delta_{\alpha \beta}}{4D^2E}&\left[-D^2(4\p_v^2E +U^2\p_r^2E +4U\p_{v}\p_{r}E) + 2C^2\p_rU\p_rE- 2CU\p_rC\p_rE+2\p_zU \p_zE\right. \notag \\
&-2D^2(\p_vU\p_rE-\p_rU\p_vE)+4\p_vC(\p_zE -C\p_rE) +2(U\p_rC-C\p_rU)\p_zE \notag \\
&\left.- 2 C\p_zU\p_rE \right], 
\end{align}

\begin{align}
R_{2021}=\frac{1}{4D^3} &\left[-2D(C\p^2_rC+DU\p^2_rD) -4D^2\p_{v}\p_{r}D +2D\p_{r}\p_{z}C\right.  \notag \\
&\left. -D(\p_rC)^2 +2\p_rC(C\p_rD -\p_zD) -2D^2\p_rU\p_rD  \right], 
\end{align}

\begin{align}
R_{2 \alpha 2 \beta}= \frac{\delta_{\alpha \beta}}{D^3E}&\left[-D(C^2\p^2_rE +\p_z^2E)+2CD\p_{r}\p_{z}E  -CD\p_rC\p_rE+(C^2-D^2U)\p_rD\p_rE  \right. \notag \\ 
&\left.+\p_zD\p_zE +(D\p_zC -C\p_zD)\p_rE  -D^2(\p_vD\p_rE+\p_rD\p_vE)-C\p_rD\p_zE \right], 
\end{align}

\begin{align}
R_{\alpha 0 \beta 1}=-\frac{\delta_{\alpha \beta}}{2D^2E} &\left[D^2(U\p^2_rE +2\p_{v}\p_{r}E +\p_rU\p_rE) +C\p_rC\p_rE -\p_rC\p_zE\right], 
\end{align}

\begin{align}
R_{\alpha 0 \beta 2} =\frac{\delta_{\alpha \beta}}{2D^2E} &\left[2D (\p_r^2 E C-\p_{r}\p_{z}E)+ (D\p_r C-2C \p_r D)\p _r E  +2 \p_r D \p_z E \right], 
\end{align}

\begin{align}
R_{\alpha 1 \beta 2}=\frac{\delta_{\alpha \beta}}{4D^2E}&\left[ 2CDU\p^2_rE +4CD\p_{v}\p_{r}E -4D\p_{v}\p_{z}E -2DU\p_{r}\p_{z}E -2D\p_rC\p_vE +(2CD\p_rU \right. \notag \\
&\left.-2D\p_zU -DU\p_rC-4C\p_vD-2CU\p_rD)\p_rE +2(2\p_vD+U\p_rD)\p_zE \right], 
\end{align}

\begin{align}
R_{\alpha \beta \gamma \delta}=\frac{\delta_{\alpha \gamma} \delta_{\beta \delta}}{D^2E^2}\left[ D^2 -(C^2+D^2U)(\p_rE)^2 -(\p_zE)^2 -2\p_rE(D^2\p_vE-C\p_zE)\right]. 
\end{align}

The non-zero components of the Ricci tensor are
\be
R_{00}= -\frac{\p_r^2 D}{D}-(d-3)\frac{\p_r^2 E}{E},
\ee
\begin{align}
R_{01}=-\frac{1}{2D^3} &\left[ D^3 \p_r^2 U +CD\p_r^2C +D^2U\p_r^2D +2D^2\p_{v}\p_{r}D -D\p_{r}\p_{z}C \right.  \notag \\
&\left. +D(\p_rC)^2-\p_rC(C \p_rD-\p_zD) +D^2\p_rU\p_rD \right] \notag \\
& \hspace{-11.3mm} -\frac{(d-3)}{2D^2E}\left[ D^2U\p^2_rE+2D^2\p_{v}\p_{r}E +\p_rC(C \p_rE-\p_zE)+D^2\p_rU\p_rE  \right],
\end{align}

\begin{align}
 R_{02} =\frac{1}{2D} &\left[  \p_r^2 C - \left( \frac{\p_r D}{D} - (d-3) \frac{\p _r E}{E} \right) \right.  \p_r C  +
2 (d-3) \left(  \frac{\p_r^2 E}{E} - \frac{\p_r D \p_r E}{DE} \right) C \notag \\ 
&\left.  +2(d-3) \left( \frac{ \p_r D \p_z E}{DE} -  \frac{\p_{r}\p_{z}E}{E} \right) \right],  
\end{align}

\begin{align}
R_{11}=\frac{1}{4D^3} &\left[2D(C^2\p_r^2U+ \p_z^2U)- 2CDU\p_r^2C -D^2(4 \p_v^2D+ U^2\p_r^2D)- 4CD\p_{r}\p_{z}U - 4D^2U \p_{v}\p_{r} D\right. \notag \\
& -2D(2C\p_{v}\p_{r}C-2\p_{v}\p_{z}C-U\p_{r}\p_{z}C)- 2C^2 \p_rU\p_rD + 2CU\p_rC \p_rD -2\p_zU\p_zD  \notag \\
&-2D^2(\p_vU\p_rD -\p_rU\p_vD) -2D(\p_rU\p_zC -\p_zU\p_rC) +2C(\p_rU\p_zD +\p_zU \p_rD) \notag \\
&\left. + 4 \p_vC (C\p_rD-\p_zD)- 2U \p_rC\p_zD  \right] \notag \\
& \hspace{-8mm}  +\frac{(d-3)}{4D^2E}\left[-D^2(4\p_v^2E +U^2\p_r^2E +4U\p_{v}\p_{r}E) + 2C^2\p_rU\p_rE- 2CU\p_rC\p_rE+2\p_zU \p_zE\right. \notag \\
&-2D^2(\p_vU\p_rE-\p_rU\p_vE)+4\p_vC(\p_zE -C\p_rE) +2(U\p_rC-C\p_rU)\p_zE \notag \\
&\left.- 2 C\p_zU\p_rE \right], 
\end{align}

\begin{align}
R_{12}=\frac{1}{4D^2} &\left[ 2CD \p^2_{r}U- DU\p^2_rC -2D\p_{r}\p_{z}U -2D\p_{v}\p_{r}C -2(C \p_rU-\p_zU)\p_rD \right.\notag \\ & \left. +4\p_vC\p_rD -2\p_rC\p_vD +U\p_rC\p_rD \right]  \notag \\
& \hspace{-8mm}  +\frac{(d-3)}{4D^2E}\left[ 2CDU\p^2_rE +4CD\p_{v}\p_{r}E -4D\p_{v}\p_{z}E -2DU\p_{r}\p_{z}E -2D\p_rC\p_vE \right. \notag \\
&\left.+(2CD\p_rU-2D\p_zU-DU\p_rC-4C\p_vD-2CU\p_rD)\p_rE +2(2\p_vD+U\p_rD)\p_zE \right], 
\end{align}

\begin{align}
R_{22}=\frac{1}{2D^3} &\left[-2D(C\p^2_rC+DU\p^2_rD) -4D^2\p_{v}\p_{r}D +2D\p_{r}\p_{z}C\right.  \notag \\
&\left. -D(\p_rC)^2 +2\p_rC(C\p_rD -\p_zD) -2D^2\p_rU\p_rD  \right] \notag \\
& \hspace{-8mm} +\frac{ (d-3)}{D^3E}\left[-D(C^2\p^2_rE +\p_z^2E)+2CD\p_{r}\p_{z}E  -CD\p_rC\p_rE+(C^2-D^2U)\p_rD\p_rE  \right. \notag \\ 
&\left.+\p_zD\p_zE +(D\p_zC -C\p_zD)\p_rE  -D^2(\p_vD\p_rE+\p_rD\p_vE)-C\p_rD\p_zE \right], 
\end{align}

\begin{align}
R_{\alpha \beta}=\delta_{\alpha \beta} \left(  \frac{1}{D^3E} \right. &\left[-D(C^2+D^2U)\p^2_rE -D\p^2_zE -2D^3\p_{v}\p_{r}E  +2CD\p_{r}\p_{z}E -D^3\p_rU\p_rE \right.  \notag \\
&-2CD\p_rC\p_rE +(C^2-D^2U)\p_rD\p_rE +\p_zD\p_zE -D^2(\p_vD\p_rE+\p_rD\p_vE) \notag \\
&\left.  +D(\p_rC\p_zE+\p_zC\p_rE) -C(\p_rD\p_zE+\p_zD\p_rE) \right] \notag \\
 & \hspace{-10mm}  \left. +\frac{(d-4)}{D^2E^2}\left[ D^2 -(C^2+D^2U)(\p_rE)^2 -(\p_zE)^2 -2\p_rE(D^2\p_vE-C\p_zE)\right] \right). 
\end{align}

\end{document}